\documentclass[sigconf]{acmart} 
\usepackage{amsmath}

\usepackage{amssymb}
\usepackage{booktabs}
\usepackage{multirow}
\usepackage{multicol}
\usepackage{mathtools}
\usepackage{amsthm}
\usepackage{xspace}
\usepackage[capitalize,noabbrev]{cleveref}
\usepackage{colortbl}
\theoremstyle{plain}

\usepackage{subcaption}
\theoremstyle{definition}

\theoremstyle{remark}

\usepackage[textsize=tiny]{todonotes}
\definecolor{mygray}{RGB}{247,247,247}
\usepackage{enumitem}
\usepackage{listings}
\usepackage{xcolor}
\usepackage[capitalize]{cleveref}
\crefname{section}{Sec.}{Secs.}
\Crefname{section}{Section}{Sections}
\Crefname{table}{Table}{Tables}
\crefname{table}{Tab.}{Tabs.}
\usepackage{tcolorbox}
\newcommand{\mybox}[1]{%
  \begin{tcolorbox}[colback=mygray,colframe=black,lowerbox=invisible,savelowerto=\jobname_ex.tex]
    \emph{#1}
  \end{tcolorbox}
}
\usepackage{url}
\usepackage{tcolorbox} 
\usepackage{listings} 
\usepackage[frozencache,cachedir=minted-cache]{minted}
\usepackage{xcolor} 
\usepackage{caption}
\usepackage{enumitem}
\usepackage{wrapfig}

\definecolor{problemcolor}{RGB}{153,0,153}
\definecolor{modelcolor}{RGB}{0, 0, 255}
\newcommand{\promptfirst}{\texttt{P\_T}\xspace}
\newcommand{\promptsecond}{\texttt{P\_T\_CC}\xspace}
\newcommand{\promptthird}{\texttt{P\_T\_IC}\xspace}
\newcommand{\promptfour}{\texttt{P\_CC}\xspace}
\newcommand{\promptfive}{\texttt{P\_IC}\xspace}

\newtcolorbox{problembox}[1][]{
    colback=problemcolor!5!white,
    colframe=problemcolor!75!black,
    title=\textbf{Problem},
    top=2mm,
    bottom=2mm,
    left=2mm,
    right=2mm,
    arc=2mm,
    boxsep=1mm,
    boxrule=0mm,
    #1
}

\newtcolorbox{modelbox}[1][]{
    colback=modelcolor!5!white,
    colframe=modelcolor!75!black,
    title=\textbf{Model},
    top=2mm,
    bottom=2mm,
    left=2mm,
    right=2mm,
    arc=2mm,
    boxsep=1mm,
    boxrule=0mm,
    #1
}

\setminted{breaklines}    
\setminted{fontsize=\scriptsize} 

\AtBeginDocument{%
  }

\begin{document}

\title{Measuring the Influence of Incorrect Code on\\
Test Generation}

\author{Dong Huang}
\email{dhuang@cs.hku.hk}
\affiliation{%
  \institution{University of Hong Kong}
\country{China}
}

\author{Jie M. Zhang}
\email{jie.zhang@kcl.ac.uk}
\affiliation{%
  \institution{King's College London}
  \country{UK}
}
\author{Mark Harman}
\email{mark.harman@ucl.ac.uk}
\affiliation{%
  \institution{University College London}
  \country{UK}
}
\author{Mingzhe Du}
\email{mingzhe@nus.edu.sg}
\affiliation{%
  \institution{National University of Singapore}
\country{Singapore}
}
\author{Heming Cui}
\email{heming@cs.hku.hk}
\affiliation{%
  \institution{University of Hong Kong}
  \country{China}
}


\keywords{Do, Not, Us, This, Code, Put, the, Correct, Terms, for,
  Your, Paper}

\received{20 February 2007}
\received[revised]{12 March 2009}
\received[accepted]{5 June 2009}


\begin{abstract}
It is natural to suppose that a Large Language Model is more likely to generate correct test cases when prompted with correct code under test, compared to incorrect code under test.
However, the size of this effect has never been previously measured, despite its obvious importance for both practicing software engineers and researchers.
To answer the question, we conducted a comprehensive empirical study on 5 open source and 6 closed source language models, with 3 widely-used benchmark data sets together with 41 repo-level real-world examples from two different real-world data sets.
Our results reveal that, when compared to incorrect code under test, LLMs prompted with correct code achieve improvements in test accuracy, code coverage, and bug detection of 57\%, 12\%, and 24\% respectively. 
We further show that these scientific conclusions carry over from the three benchmark data sets to the real-world code, where tests generated for incorrect code experience a 47\% worse bug detection rate.
Finally, we report that improvements of +18\% in accuracy, +4\% coverage, and +34\% in bug detection can be achieved by providing natural language code descriptions.
These findings have actionable conclusions. 
For example, the 47\% reduction in real-world bug detection is a clear concern.
Fortunately, it is a concern for which our findings about the added value of descriptions offer an immediately actionable remedy.

\end{abstract}




\received{20 February 2007}
\received[revised]{12 March 2009}
\received[accepted]{5 June 2009}

\maketitle

\section{Introduction}

Automatic test case generation is an an increasingly important part of the software development process, enriching the effectiveness of test cases and ensuring that the software under development adheres to the specified requirements and operates as intended~\cite{wang2024testeval,arcuri2018experience}. Recently, many research works have harnessed the capabilities of large language models (LLMs) to generate test cases automatically\cite{chen2022codet,huang2023codecot,huang2023agentcoder,olausson2023self,schafer2023empirical,zhong2024ldb,hong2023metagpt,yuan2024evaluating,yang2024enhancing,yang2023white,deng2024large,deng2023large}. The information provided with LLMs typically includes two aspects: the source code under test and/or the code's task description. For example, FuzzGPT \cite{deng2024large}, TitanFuzz \cite{deng2023large}, KernelGPT \cite{yang2023kernelgpt}, and CodaMOSA \cite{LemieuxILS23} provide LLMs with the source code under test only for LLMs to generate tests automatically. CodeCoT \cite{huang2023codecot} uses both the task description and the source code under test. AgentCoder \cite{huang2023agentcoder} and MetaGPT \cite{hong2023metagpt} directly provide the task description to LLMs without the source code.

Although generating tests with LLMs based on the source code under test is a common practice, it poses a significant challenge that is often overlooked. Specifically, if the source code under test contains bugs, the tests generated by LLMs may inherit flawed logic or assumptions from the code, resulting in ineffective or incorrect tests. The relationship between the correctness of the source code and the effectiveness of the generated test cases, however, remains largely unexplored.

To fill this gap, in this paper, we present the first systematic empirical study on how the correctness of the code under test impacts the effectiveness of the LLM-generated test cases. We evaluate the effectiveness of test cases by measuring their accuracy\footnote{Both ``accuracy'' and ``correctness'' are widely used in the literature to refer to the ratio of the test cases that pass correct code against the total number of generated test cases \cite{wang2024testeval,yuan2024evaluating,liao2024llms,jin2024generating,li2024large,li2023prompting,chen2022codet}. We use the term accuracy in our paper.} and coverage in the correct code provided by the evaluated dataset. We also evaluate their bug detection ratio using our collected bug set.

We first conduct experiments using 5 open-source and 6 closed-source LLMs on three widely-studied code generation datasets (i.e., HumanEval \cite{niu2024evaluating}, MBPP \cite{Austin2021}, and APPS \cite{HendrycksBKMAGB21}). For each code generation task, we prompt each LLM to generate test cases based on five different prompts: (1) task description only, (2) task description with correct code, (3) task description with incorrect code, (4) correct code only, and (5) incorrect code only. We then evaluate the effectiveness of LLM-generated test cases in three dimensions: accuracy, coverage, and bug detection ratio. We also examine whether LLMs are more prone to being misled by the code they generate themselves. Moreover, we evaluate LLMs with incorrect code from BugsInPy \cite{widyasari2020bugsinpy} and SWE-Bench \cite{jimenez2023swe}, two datasets comprising tasks extracted from real-world GitHub commits to check whether our observations hold for real-world scenarios.

Our results demonstrate that incorrect code under test can significantly impact the ability of LLMs to generate effective tests. For example, on the HumanEval dataset, test cases generated by LLMs achieve an accuracy of 80.4\%, a coverage of 98.4\%, and a bug detection ratio of 87.4\% when both the task descriptions and the correct code under test are included in the prompt. However, when the code under test is incorrect, these results drop to 57.1\%, 91.7\%, and 75.0\%, respectively. We also observe that LLMs are less likely to be misguided by the code they generate by themselves. Finally, our experiments with real-world tasks demonstrate the same conclusions as those of widely adopted benchmarks, although the accuracy, coverage, and bug detection ratio are much lower than those on the three simpler benchmarks. In particular, for the bug detection ratio, LLMs with correct code under test detect 13.5\% of the bugs on average, but LLMs with incorrect code under test detect only 7.1\% on average.

In conclusion, this paper makes the following contributions:
\begin{itemize}

    \item We present the first systematic study on the influence of source code on test case generation. 
    \item Our evaluation results demonstrate that providing task descriptions with correct code yields higher performance in test case generation compared to using other prompts. For instance, in the HumanEval dataset, LLM-generated test cases achieve an accuracy of 80.4\% for all models when task descriptions and correct code are provided. Conversely, when provided with task descriptions and incorrect code, the average accuracy declines substantially to 57.1\%. 
    \item We provide implications for developers and researchers on using LLMs for generating tests automatically based on our observations. In particular, our finding indicates that \textbf{LLM-based testing will be more effective at generating tests to protect mature code from regression errors. However, if used in the early stage of software development on relatively immature code, it will be more likely to ``bake in'' errors.} 
    We also call for more research to improve LLMs’ resilience against incorrect code in generating reliable and bug-revealing tests. 
\end{itemize}

\section{Background and Related Work}

\subsection{LLMs for Source Code Generation}
LLMs have seen a boost in adoption in code generation, driven by the availability of extensive open-source code repositories and the demand for enhanced developer productivity. Pioneering works have exclusively focused on generating functionally correct code from natural language instructions, including CodeT5~\cite{WangCodeT52021}, AlphaCode~\cite{Lialphacode2022}, CodeGen~\cite{NijkampPHTWZSX23}, InCoder~\cite{FriedAL0WSZYZL23}, StarCoder~\cite{LiStarCoder203}, SantaCoder~\cite{Loubnasanta2023}, and DeepSeek Coder~\cite{DeepSeekcoder}. With the rapid scale expansion of LLMs, subsequent advancements have produced models such as Codex~\cite{ChenCodex2021} and CodeLLaMA~\cite{Roziere2023}. These models are fine-tuned from foundational LLMs~\cite{BrownMRSKDNSSAA20, Touvron2023} and are proficient in a variety of tasks, including code generation~\cite{ChenCodex2021, Dai2024mhpp, huang2024effibench,huang2024effi, du2024mercury}, program repair~\cite{Haque2022, JiangLLT23}, automated testing~\cite{LemieuxILS23, Deng2023}, type prediction~\cite{MirLPG22, WeiDD23}, and code summarization~\cite{HasanMIMHHAIS21, AhmedD22}. Among these, model performance on the code generation task has emerged as a pivotal benchmark for evaluating the holistic coding capability of LLMs.

To enhance the functional correctness of generated source code, feedback-based refinement techniques have been employed. These methods mimic the human learning process, where individuals enhance their knowledge through trial and error~\cite{boyd1983reflective, metcalfe2017learning}. Initial efforts revolved around human feedback for model evaluation and refinement~\cite{KreutzerKMR18, Ouyang0JAWMZASR22}. To reduce human intervention, automated feedback approaches have been explored, utilizing signals from various sources, including LLM self-reflection~\cite{MadaanSelfRefine23, huang2023bias}, dedicated verification models~\cite{Lu2024autocv}, external tools~\cite{huang2023agentcoder, huang2023codecot}, and external knowledge sources~\cite{GaoRARR23}. For example, Self-Evolve~\cite{JiangSelfEvolve23} and EffiLearner \cite{huang2024effilearner} execute the initially generated program on canonical test cases and provide the execution results as feedback to prompt the LLM to refine the code. Furthermore, Self-Debug~\cite{ChenSelfDebug23} incorporates multiple feedback sources, including program explanations, unit tests, and program interpreters. Notably, ALGO~\cite{ZhangAlgo23} takes a more detailed approach to generate a reference oracle program via an exhaustive search.

\subsection{Improving Source Code with Tests}
In the current code evaluation paradigm~\cite{huang2023codecot, hong2023metagpt,wang2023review,sun2024survey}, an LLM starts by tentatively generating source code based on the given task description and then validating the functionality of the code through a set of pre-defined test cases. These test cases are executed and are expected to identify any code errors and inconsistencies between the generated code and the given task description. Consequently, developing appropriate test cases is vital for accurately assessing code generation tasks. However, highly effective public test cases are not always available.
To address this, researchers have harnessed LLMs to generate test cases~\cite{chen2022codet, huang2023codecot, huang2023agentcoder, olausson2023self, schafer2023empirical, zhong2024ldb, hong2023metagpt}. Tools like CodeT~\cite{chen2022codet} generate test cases directly for the source code, minimizing human effort and expanding test scenario coverage. CodeChain~\cite{le2023codechain} enhances this by devising prompt templates to format the generated test cases. 
CodeArena~\cite{du2025codearena} synergizes multiple LLMs to generate more robust and reliable test cases. CodeCoT~\cite{huang2023codecot} advances further by generating both source code and test cases simultaneously. AgentCoder \cite{huang2023agentcoder} and MetaGPT \cite{hong2023metagpt} decompose the software development process into multiple stages, with each stage managed by specialized agents. Test designer agents, for example, are proficient in generating reliable test cases based on the task description.

\subsection{Improving Effectiveness of Test Generation}
Low-effectiveness test cases can mislead the debugging process, resulting in incorrect conclusions and suboptimal code refinement~\cite{chen2022codet, huang2023agentcoder, hong2023metagpt}.
One potential issue arises when the generated test cases are misaligned with the given problem instructions. In the code debugging process, even if the generated code is correct, it may fail to pass erroneous tests, leading the LLM to unnecessarily rectify the code and potentially introduce new errors. Similarly, in software testing, the developed software may raise errors when incorrect test cases are used to analyze its correctness. The errors raised by incorrect code may also cause developers to revise the source code and introduce new errors. Another concern is the coverage of the generated test cases \cite{chen2022codet,huang2023codecot}. If the test cases only cover a limited range of common behaviors and fail to account for edge cases or specific task requirements, the generated code may pass all tests while still being incomplete or incorrect. This can give a false sense of confidence in the code's correctness, as it has not been thoroughly validated against all relevant scenarios. To enhance test case effectiveness, several prompt engineering techniques are employed, which involve using source code-guided and non-source code-guided approaches. Frameworks like CodeT~\cite{chen2022codet}, AgentCoder~\cite{huang2023agentcoder}, MetaGPT~\cite{hong2023metagpt}, LATS~\cite{zhou2023language}, and Reflexion~\cite{shinn2024reflexion} generate test cases based solely on task descriptions. In contrast, CodeCoT~\cite{huang2023codecot}, ATHENATEST~\cite{tufano2020unit}, EvalPlus~\cite{liu2024your}, and CodaMOSA~\cite{LemieuxILS23} leverage existing source code to generate test cases. Though these methods show promise, the impact of incorporating source code on test case effectiveness is not comprehensively understood. This paper aims to empirically study whether source code inclusion consistently enhances the effectiveness of LLM-generated test cases, compared to using task descriptions alone.


\section{Method}
\label{sec:method}
This section introduces our method for generating, extracting, and executing tests, as well as our measurements of test effectiveness. 

\begin{table}
    \centering
    \vspace{-1em}
    \caption{The five prompts used in our empirical study for generating test cases with LLMs.}
    \vspace{-2mm}
     \resizebox{0.7\linewidth}{!}{
    \begin{tabular}{l|l}
    \toprule
    Prompt&Template\\
    \midrule
         \promptfirst&\textbf{T}ask description\\
         \promptsecond&\textbf{T}ask description + \textbf{C}orrect \textbf{C}ode \\
         \promptthird&\textbf{T}ask description + \textbf{I}ncorrect \textbf{C}ode\\
         \promptfour&\textbf{C}orrect \textbf{C}ode\\
         \promptfive&\textbf{I}ncorrect \textbf{C}ode\\
         \midrule
    \end{tabular}}
    \vspace{-1em}
    \label{tab:template}
\end{table}

\subsection{Prompt Construction} 

The first step in our study is prompt construction. In our experiments, we have five prompts for each task that requires LLMs to generate code (See \cref{tab:template}). The first prompt (\promptfirst) is the \textbf{T}ask description. For this prompt, we follow the setup of existing works \cite{GPT4, ChenCodex2021}, and directly ask LLMs to generate test cases for each task based on the task description with zero-shot prompting. The second prompt (\promptsecond) in our experiments is \textbf{T}ask description + \textbf{C}orrect code. For the HumanEval, MBPP, and APPS datasets, we directly use the correct code provided by each dataset to represent the correct code in our experiments. For BugsInPy and SWE-Bench, we use the patched code as the correct code in our experiments. The third prompt (\promptthird) in our experiments is the \textbf{T}ask description + \textbf{I}ncorrect code. For the incorrect code, we first require LLMs evaluated in our experiments to generate code with zero-shot prompting for the HumanEval, MBPP, and APPS datasets, and then collect incorrect pieces of code for each task in our evaluated dataset and then randomly select an incorrect code that will be used in all models as the \promptthird's incorrect code part. For BugsInPy and SWE-Bench, we directly use the pre-patch source code as the incorrect code. For the fourth prompt (\promptfour), we use \textbf{C}orrect code without task description. The fifth prompt (\promptfive) is directly used as an \textbf{I}ncorrect solution without a task description.
In our experiments, the correct source code for \promptsecond and \promptfour is the same for each task, and the incorrect source code for \promptthird and \promptfive is also the same for each task.

Finally, to ensure that the test cases generated by LLMs follow the test case format rather than pure natural language in the experiments, we also provide the test case template \texttt{assert function\_name(input\_parameters) == output} before the task description so that the test cases generated by LLMs can follow the same format and be directly used in our experiments.

\subsection{Tests Extraction and Script Writing} 

To ensure that the test cases can be extracted from the LLMs' response, we constrain LLMs to generate test cases in the \texttt{\`}\texttt{\`}\texttt{\`} python[test\_case] \texttt{\`}\texttt{\`}\texttt{\`} so that we can directly extract test cases from \texttt{\`}\texttt{\`}\texttt{\`}python and \texttt{\`}\texttt{\`}\texttt{\`} that can remove the natural language in the test cases\footnote{This section (Section~\ref{sec:method}) by default introduces the configuration of python related tasks. We explore whether the observations hold for other languages in \cref{sec:multilanguage}.}\footnote{Sometimes LLMs generate test cases with some natural language explanations \cite{hong2023metagpt,huang2023codecot}.}. After extracting tests from the LLM-generated response, we use the HumanEval-provided script to automatically write the source code (e.g., correct code for the accuracy and coverage evaluation) and the LLM-generated tests into the script. For the required libraries for each task, we directly import them based on the dataset (e.g., HumanEval) setup, which avoids errors caused by the script's lack of necessary libraries in the experiments.

\subsection{Source Code Execution} 


For accuracy and coverage, we conduct experiments on the correct code provided by each dataset. For bug detection experiments, we execute LLM-generated test cases using the constructed bug detection source code. During the code execution process, we set the timeout value to 5 seconds for all tasks to ensure the code can be executed with all test cases and does not require much time. To speed up the testing process, we use concurrency in our accuracy and bug detection experiments and set the maximum number of workers to 20, which can reduce the overhead of the testing process. Since we employ the \texttt{coverage.py} library~\footnote{\texttt{coverage.py} Library: \url{https://github.com/nedbat/coveragepy}} for the coverage experiments, which cannot support the concurrency setting, we opt to execute all tests using a single-threaded script instead.

\subsection{Effectiveness Measurement}

We evaluate the effectiveness of LLM-generated test cases using three primary metrics: (1) the accuracy of LLM-generated test cases (Accuracy), (2) code line coverage of LLM-generated test cases in the correct code (Coverage), and (3) bug detection effectiveness of LLM-generated test cases (Bug Detection). Additionally, to assess the consistency and quality of LLM-generated test cases, we employ CodeBLEU to measure the similarity of tests generated by the LLM at different time points.

\subsubsection{Accuracy}

We assess the accuracy of LLM-generated test cases by computing the number of test cases generated by LLMs that successfully pass the correct code provided by the dataset\footnote{For the HumanEval, MBPP, and APPS datasets, we use the ``canonical solution'' provided by the dataset as the correct code in our experiments. For BugsInPy and SWE-Bench, we employ the patched code as the correct code.}. A test case generated by an LLM is considered correct if it passes the correct code, i.e., when the input of the test case is fed into the correct code, the output matches the expected output of the test case. We analyze effectiveness at two levels in our experiments: test level and task level. At the \textbf{test level}, we analyze the accuracy of LLM-generated test cases for each task individually. For example, if GPT-3.5-turbo generates test cases for Task 1 in HumanEval consisting of ten test cases, and seven of these test cases are correct while three test cases are incorrect, the test level accuracy would be calculated as 70\% (7/10) for Task 1. At the \textbf{task level}, we consider LLM-generated test cases to be correct only if all test cases successfully pass the correct code. 
In the previous example, we treated LLM-generated test cases are incorrect code as three of the test cases of GPT-3.5-turbo are incorrect.

\subsubsection{Coverage}
We use the \texttt{coverage.py} package to calculate the line-level coverage of the test cases on the correct code provided by the dataset. To calculate the coverage of LLM-generated test cases, we consider two different scenarios based on the accuracy result, i.e., coverage for correct tests at the test level and coverage for correct tests at the task level. 
The former measures the percentage of code lines in the correct code executed by all correct tests at the test level. 
The latter measures the percentage of code lines in the correct code executed by correct tests at the task level.

\subsubsection{Bug Detection}

To measure the bug detection efficacy of the LLM-generated test cases, we first construct a bug set for each dataset (more details in \cref{sec:benchmark}). We then analyze whether the LLM-generated test cases can discover bugs in our constructed bug set. Similar to the coverage measurement, we consider two different scenarios: (1) bug detection for correct tests at the test level and (2) bug detection for correct tests at the task level. Bug detection for correct tests at the test level measures the percentage of bug code in our constructed code detected by LLM-generated correct tests at the test level. Bug detection for correct tests at the task level measures the percentage of bugs in our constructed code solutions that can be detected by the correct test cases at the task level.

\subsubsection{CodeBLEU}
To measure the consistency and quality of LLM-generated code and test cases, we also use CodeBLEU\footnote{\url{https://pypi.org/project/codebleu/}} \cite{ren2020codebleu}, a metric specifically designed for evaluating code-related similarity. This metric allows us to quantify the similarity between different test cases or code generated for the same tasks.

Note that we do not adopt mutation testing as a measurement considering that our bug set is much more realistic than mutants, as the bugs have been representative of the actual errors produced by LLMs. In contrast, mutation testing involves making small syntactic changes (i.e., introducing artificial bugs) in the correct code, which can have a big difference with real bugs in both semantics and syntactics~\cite{papadakis2019mutation}.

\section{Experiment Design}\label{sec:experiment}

\subsection{Research Questions}\label{sec:rq}
This study answers the following questions:

    \noindent \textbf{RQ1: How does the source code in prompts affect LLMs in test generation?} This RQ investigates the effectiveness of LLM-generate test cases in terms of test case accuracy, coverage, and bug detection effectiveness among the five test case generation prompts. There are three sub-RQs:
            \begin{itemize}
                \item \emph{RQ1.1 What is the \textbf{accuracy} of LLM-generated test cases for the five different prompts?}
                \item \emph{RQ1.2: What is the code \textbf{coverage} of LLM-generated test cases for the five different prompts?}
                \item \emph{RQ1.3: What is the \textbf{bug detection effectiveness} of LLM-generated test cases in our constructed pieces for the five different prompts?}

            \end{itemize}
    \noindent \textbf{RQ2: How does the source of the code influence the LLMs in test generation?} This RQ investigates whether LLM-generated tests are more likely misguided by LLM-generated code rather than our constructed \promptsecond and \promptthird.\\
    \noindent \textbf{RQ3: To what extent are LLMs misguided by the incorrect code in test generation?} This RQ analyzes the percentage of LLM-generated test cases that can pass the incorrect code in \promptfive.\\
    \noindent \textbf{RQ4: How does code incorrectness degree impact test case generation?} This RQ investigates the effectiveness of LLM-generated test cases based on source code with different levels of deviation from the correct implementation.\\
    \noindent \textbf{RQ5: Do our observations hold for real-world code?} This RQ investigates the effectiveness of LLM-generated test cases based on the source code of real-world tasks.

\begin{table}
    \centering
    \caption{Code generation datasets used in the experiments. The tokens are calculated based on tiktoken with GPT-4. }
    \vspace{-3mm}
    \resizebox{\linewidth}{!}{
    \begin{tabular}{l|rrrrrrr}
    \toprule
    Dataset&Mean& Mean& Mean& Mean& Mean& No. of&No. of\\
         &Token&Token& Token& Token& Token& Problems&Bug code\\
         & \promptfirst&\promptsecond&\promptthird&\promptfour&\promptfive&&\\
         \midrule
HumanEval&117.2&164.2&198.7&58.9&59.8&85&85\\
MBPP&123.0&162.8&191.5&51.0&55.9&213&213\\
APPS&486.3&571.1&541.7&94.8&56.1&172&172\\
BugsInPy&-&-&-&1092.2&904.0&10&10\\
SWEBench&-&-&-&2529.3&2498.4&31&31\\
         \bottomrule
    \end{tabular}
    }
    \label{tab:benchmarks}
    \vspace{-3mm}
\end{table}

\subsection{Datasets}\label{sec:benchmark}

We provide the detailed information of our evaluated datasets in \cref{tab:benchmarks}.
In our experiments, we first use \cite{ChenCodex2021}, MBPP \cite{austin2021program}, and APPS \cite{hendrycksapps2021} datasets, which are widely used in LLM-based code generation \cite{huang2023agentcoder,chen2022codet,hong2023metagpt,zhong2024ldb,le2023codechain} and LLM-based test case generation \cite{chen2022codet,lahiri2022interactive,fan2018oracle,ChenSelfDebug23}, to measure the effectiveness of LLM-generated tests under different prompt instructions.
To facilitate a consistent evaluation of test case generation effectiveness across datasets, we convert the prompt format of APPS and MBPP into HumanEval's function-level format for both task description and solutions, which is easier to evaluate compared to the line-level code script \cite{mathews2024test}. This conversion constrains the LLMs to generate test cases in a standardized unit test case format, simplifying the evaluation process of the generated test cases.
Next, we evaluate the effectiveness of the generated test cases on the BugsInPy \cite{widyasari2020bugsinpy} and SWE-Bench \cite{jimenez2023swe} datasets, which contain real-world Python programs with known bugs, allowing us to analyze how the source code of real-world programs affects the performance of LLM-generated test cases in detecting bugs.


HumanEval \cite{ChenCodex2021} originally comprised 164 tasks and employs the $pass@k$ metric to evaluate LLM code generation efficacy. In our experiments, some tasks are correctly addressed by all LLMs, which then do not contain incorrect code as \promptthird in our setup. Then, our analysis refined this to 85 tasks, excluding those universally solved by all LLMs. MBPP \cite{austin2021program} initially contained 974 tasks. We employed 213 tasks from the MBPP-EvalPlus version \cite{liu2024your}, adapting them to the HumanEval function format and excluding those universally solved by all LLMs. APPS originally encompassed 5,000 tasks across three difficulty levels. Prior to conducting the experiments, we first convert the task descriptions into the HumanEval format. Since the correct code provided by APPS \cite{hendrycksapps2021} is not at the function level, we use GPT-3.5-turbo to convert the correct code into function-level code, filter out incorrect converted functions, and incorporate the original task prompt into the function. After this process, we collect 405 tasks for our experiments. Next, we feed the converted tasks into the evaluated LLMs to generate solutions. For each task, we select an incorrect code from the generated solutions to construct \promptthird and \promptfive. However, since some tasks do not have incorrect code, we then only collect 172 tasks for our experiments.

BugsInPy \cite{widyasari2020bugsinpy} contains 493 real bugs from 17 real-world Python programs, including popular libraries such as matplotlib, numpy, pandas, and fastapi. Since tasks in BugsInPy does not exist a predefined task descriptions, we directly use the patched and original code as \promptfour and \promptfive, respectively. 

SWE-Bench \cite{jimenez2023swe} is a benchmark for evaluating LLMs on real-world software engineering tasks. It contains 2,294 software engineering problems drawn from real GitHub issues and corresponding pull requests across 12 popular Python repositories. Unlike synthetic datasets, SWE-Bench tasks require understanding and modifying existing codebases, making them more representative of real-world software development scenarios.

We conduct experiments on all 493 tasks and 500 tasks from the BugsInPy and SWE-bench datasets. However, we observe that for most tasks in both datasets, the generated tests for both prompts are incorrect at both the test and task levels. The primary reason is the large input token count (e.g., BugsInPy averages over 13,000 tokens), which impairs LLMs' reasoning ability and hinders useful test case generation \cite{levy2024same}. Consequently, we focus on the 10 BugsInPy and 31 SWE-Bench tasks that yield correct test cases from either correct or incorrect code, as these are most relevant for investigating the influence of code correctness on test generation.

\emph{Our constructed bug set} To measure the bug detection effectiveness of LLM-generated test cases, we construct a bug set with incorrect solutions from the tasks in each dataset. For HumanEval, MBPP, and APPS, we first require our evaluated LLMs to generate code for each task description. Then, we randomly select an incorrect code for each task to construct the bug set. Since some tasks do not have incorrect code, we filter out these tasks during the dataset construction process. Finally, we obtain 85, 213, and 172 incorrect code samples for HumanEval, MBPP, and APPS, respectively. For BugsInPy and SWE-Bench, we directly use the original incorrect code as the bug code.

\subsection{Evaluation LLMs}
Five open-source LLMs and six closed-source LLMs are used in our experiments. The experiments are conducted on an 8 * H100 server.

\subsubsection{Open-Source Models} For open-source LLMs, we evaluate \textbf{Meta-Llama-3-8B (Llama3), CodeLlama-7B-Python-hf (CodeLlama), DeepSeek-Coder-6.7B-Instruct (DeepSeek), StarCoder2-7B (StarCoder), and Codestral-22B-v0.1 (Codestral)} in our experiments. We select these open-source LLMs since they achieve SOTA performance in code generation tasks (e.g., evalplus) with low parameters. 


\subsubsection{Closed-Source Models} 
We conducted an evaluation of six closed-source LLMs: \textbf{GPT-3.5-turbo (GPT3.5), GPT-3.5-turbo-1106 (GPT3.5-1106), GPT-4-turbo (GPT4-turbo), GPT-4, Claude-3-haiku (Claude3H), and Claude-3-sonnet (Claude3S)}. These models exemplify the latest advancements in LLM architecture\footnote{GPT-3.5-turbo and GPT-3.5-turbo-1106 are variants of the GPT-3.5 series. ``GPT-3.5-turbo-1106'' indicates a release date of June 11, 2023, whereas ``GPT-3.5-turbo'' refers to a more recent iteration released on January 25, 2024.}. GPT-4-turbo and GPT-4 are the latest iterations of the GPT series, offering even more advanced capabilities compared to their predecessors. Claude-3-haiku and Claude-3-sonnet are two versions of the Claude-3 model developed by Anthropic. They have also exhibited competitive performance in code-related tasks. 


\subsection{Inference Configuration of LLMs}
In our experiments, four parameters affect the LLM response: Temperature, Top-p, Top-K, and max\_new\_tokens. To ensure consistency in the test cases generated by LLMs across different executions, we set Temperature to 0, Top-p to 1.0, Top-K to 0, and max\_new\_tokens to 1024. These settings guarantee that the generation process follows a greedy decoding approach.

\begin{table*}
    \centering
    \caption{RQ1.1: Accuracy of LLM-generated test cases across HumanEval, MBPP, and APPS. Each cell presents test-level / task-level results.  
    Notably, \promptfirst and \promptsecond consistently yield higher performance, indicating that incorrect code affects LLMs in generating accurate test cases.
    }
    \vspace{-3mm}
    \resizebox{\textwidth}{!}{
    \begin{tabular}{l|rrrrr|rrrrr|rrrrr}
    \toprule
\multirow{2}{*}{Model}&\multicolumn{5}{c}{HumanEval}&\multicolumn{5}{c}{MBPP}&\multicolumn{5}{c}{APPS} \\
 & \promptfirst & \promptsecond & \promptthird&\promptfour&\promptfive & \promptfirst & \promptsecond & \promptthird&\promptfour&\promptfive & \promptfirst & \promptsecond & \promptthird&\promptfour&\promptfive\\
 \midrule
Llama3 & 68.5 / \textbf{47.1} & \textbf{74.5} / 43.5 & 26.9 / 23.5 & 70.2 / 40.0 & 26.9 / 23.5 & \textbf{75.2} / \textbf{62.9} & 69.9 / 49.8 & 25.9 / 28.2 & 54.6 / 38.5 & 25.9 / 28.2 & \textbf{66.6} / \textbf{36.6} & 62.0 / 35.5 & 28.4 / 22.1 & 58.0 / 33.1 & 28.4 / 22.1 \\
CodeLlama & 65.1 / \textbf{43.5} & \textbf{65.8} / 31.8 & 34.3 / 16.5 & 47.4 / 8.2 & 40.5 / 11.8 & 57.2 / 59.6 & \textbf{78.7} / \textbf{87.8} & 33.0 / 21.6 & 45.3 / 41.3 & 30.2 / 17.4 & \textbf{90.6} / 61.6 & 74.5 / \textbf{80.2} & 25.0 / 20.4 & 31.1 / 18.0 & 25.1 / 19.2 \\
DeepSeek & \textbf{79.8} / \textbf{48.2} & 71.8 / 38.8 & 54.9 / 20.0 & 59.6 / 14.1 & 58.0 / 23.5 & \textbf{74.2} / \textbf{58.2} & 62.3 / 11.3 & 47.9 / 16.9 & 56.8 / 7.0 & 46.5 / 21.6 & \textbf{72.3} / \textbf{32.0} & 53.2 / 9.9 & 29.1 / 6.4 & 41.2 / 4.1 & 30.3 / 6.4 \\
StarCoder & \textbf{76.9} / \textbf{72.9} & 74.5 / 70.6 & 40.8 / 14.1 & 49.0 / 10.6 & 38.5 / 10.6 & \textbf{70.7} / \textbf{77.9} & 57.4 / 64.8 & 40.6 / 12.7 & 52.2 / 16.0 & 39.3 / 12.2 & \textbf{76.3} / \textbf{69.2} & 65.7 / 45.4 & 29.8 / 4.7 & 35.2 / 9.3 & 29.1 / 3.5 \\
Codestral & 75.5 / 38.8 & \textbf{82.5} / 35.3 & 61.3 / 36.5 & 77.1 / \textbf{40.0} & 65.0 / 35.3 & 74.6 / 41.8 & 71.9 / 31.9 & 47.8 / 42.7 & \textbf{78.4} / \textbf{49.3} & 46.9 / 26.8 & 47.3 / 11.6 & \textbf{60.4} / 13.9 & 36.8 / \textbf{40.7} & 52.7 / 19.2 & 36.1 / 38.4 \\
GPT3.5 & 80.0 / \textbf{49.4} & \textbf{82.4} / \textbf{49.4} & 62.0 / 37.6 & 32.1 / 16.5 & 58.3 / 37.6 & 75.1 / 54.0 & \textbf{81.1} / \textbf{60.6} & 46.3 / 27.2 & 77.0 / 57.8 & 51.2 / 36.1 & 56.8 / 25.6 & \textbf{61.6} / \textbf{32.0} & 33.6 / 10.5 & 52.8 / 23.3 & 34.7 / 11.1 \\
GPT3.5-1106 & 78.0 / 51.8 & \textbf{85.2} / \textbf{60.0} & 64.9 / 41.2 & 33.7 / 17.6 & 58.8 / 40.0 & 76.5 / 58.2 & \textbf{83.1} / \textbf{64.3} & 45.9 / 27.2 & 77.8 / 58.2 & 52.1 / 31.9 & 60.3 / 27.3 & \textbf{69.3} / \textbf{42.4} & 35.0 / 9.3 & 57.0 / 25.0 & 33.6 / 11.1 \\
GPT4-turbo & 87.2 / 49.4 & \textbf{89.9} / \textbf{60.0} & 72.8 / 36.5 & 86.6 / 51.8 & 72.4 / 48.2 & 82.3 / 52.1 & \textbf{86.7} / \textbf{59.1} & 65.4 / 39.0 & 85.8 / 56.8 & 57.7 / 50.7 & 68.4 / 26.7 & 73.5 / 32.0 & 50.5 / 22.1 & \textbf{75.4} / \textbf{33.7} & 56.6 / 19.8 \\
GPT4 & 82.3 / \textbf{62.4} & \textbf{88.9} / 57.6 & 76.2 / 35.3 & 84.8 / 52.9 & 69.6 / 42.4 & 77.4 / 54.9 & \textbf{86.5} / 60.1 & 64.9 / 34.7 & 77.1 / \textbf{61.5} & 57.4 / 41.8 & 65.4 / 37.8 & \textbf{74.6} / \textbf{40.1} & 53.6 / 16.9 & 65.2 / 35.5 & 51.4 / 18.0 \\
Claude3S & 76.2 / 37.6 & \textbf{83.5} / 38.8 & 63.4 / 29.4 & 72.0 / \textbf{40.0} & 63.3 / 30.6 & 69.3 / 35.7 & \textbf{78.8} / \textbf{38.5} & 49.2 / 38.0 & 71.8 / 32.9 & 47.4 / 25.8 & 53.5 / 12.8 & \textbf{63.2} / 13.4 & 37.0 / \textbf{31.4} & 54.2 / 18.6 & 35.8 / 23.8 \\
Claude3H & 91.6 / \textbf{74.1} & 85.8 / 50.6 & 71.0 / 54.1 & \textbf{92.1} / 65.9 & 78.1 / 57.6 & 57.9 / \textbf{69.5} & \textbf{73.5} / 44.6 & 42.9 / 46.5 & 66.2 / 63.9 & 52.5 / 47.9 & 80.9 / 57.0 & 67.8 / 27.3 & 58.5 / 41.9 & \textbf{85.2} / \textbf{65.7} & 52.8 / 34.3 \\
\midrule
Overall & 78.3 / \textbf{52.3} & \textbf{80.4} / 48.8 & 57.1 / 31.3 & 64.1 / 32.5 & 57.2 / 32.8 & 71.9 / \textbf{56.8} & \textbf{75.4} / 52.1 & 46.3 / 30.4 & 67.5 / 43.9 & 46.1 / 30.9 & \textbf{67.1} / \textbf{36.2} & 66.0 / 33.8 & 37.9 / 20.6 & 55.3 / 26.0 & 37.6 / 18.9 \\

\bottomrule
    \end{tabular}
    }
    \label{tab:correctness}
    \vspace{-2mm}
\end{table*}

\section{Results and Findings}

\subsection{RQ1: How does the source code in prompts affect LLMs in test generation?}

\subsubsection{RQ1.1: \textbf{Accuracy} of LLM-generated test cases}

\begin{figure}
    \centering
    \begin{subfigure}[b]{1\linewidth}
        \centering
        \includegraphics[width=1\linewidth]{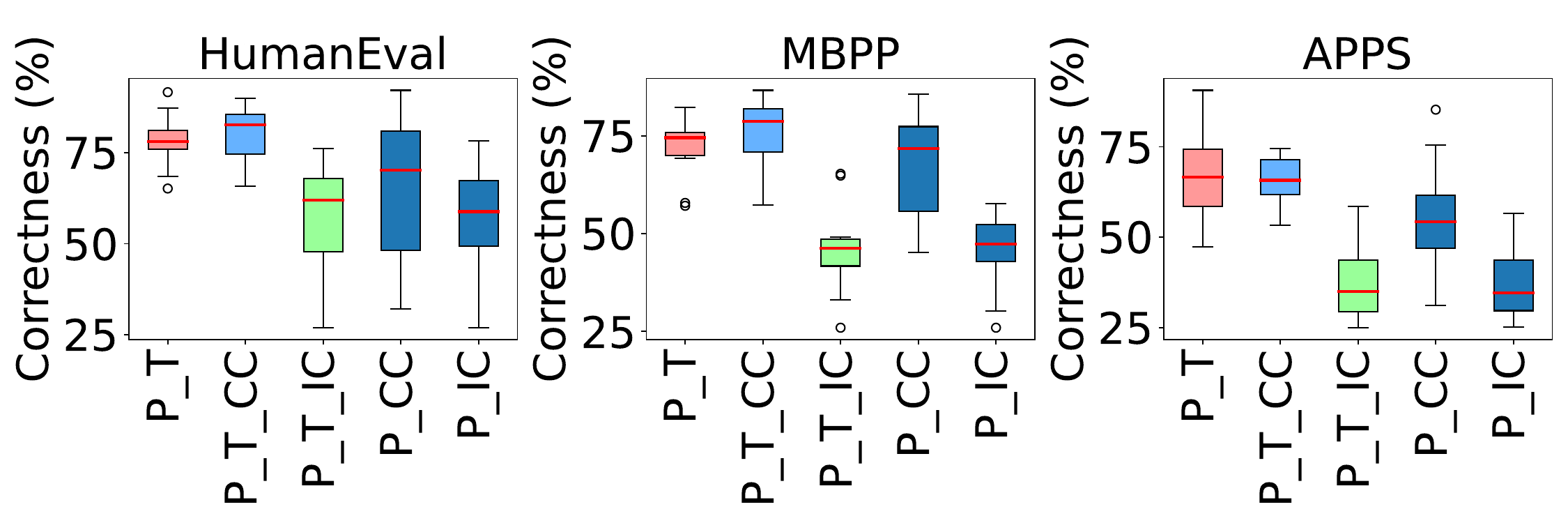}
        \vspace{-5mm}
        \caption{Test level.}
        \label{fig:rq11a}
    \end{subfigure}
    \begin{subfigure}[b]{1\linewidth}
        \centering
        \includegraphics[width=1\linewidth]{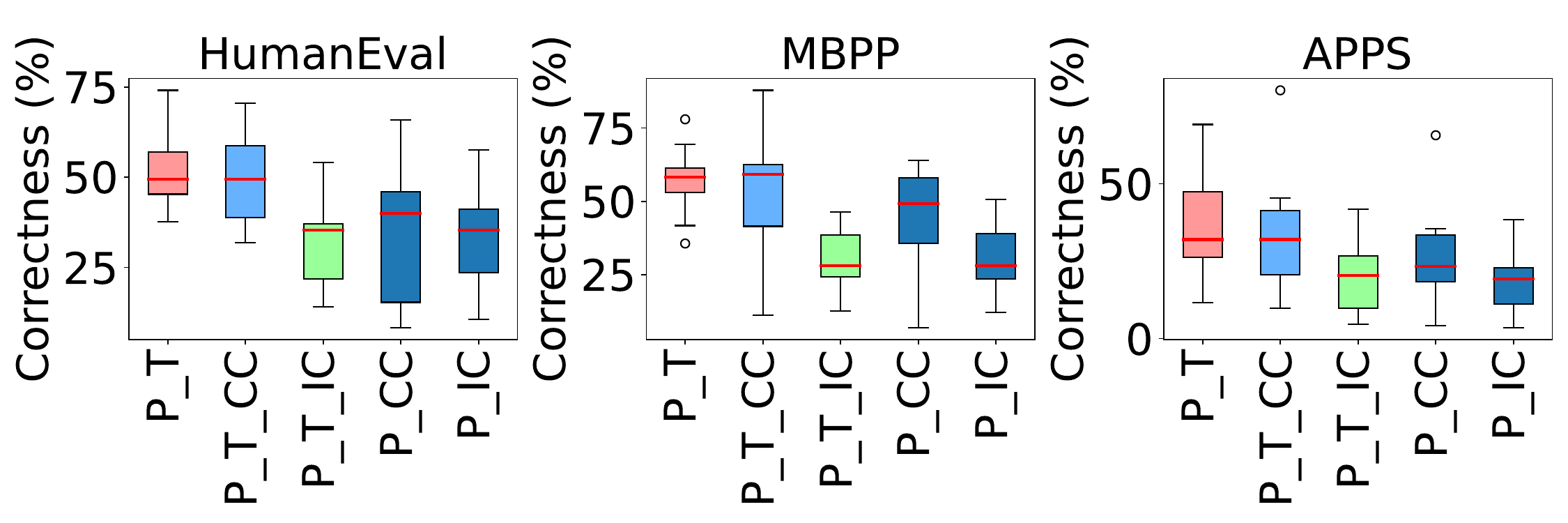}
        \vspace{-5mm}
        \caption{Task level.}
        \label{fig:rq11b}
    \end{subfigure}
    \vspace{-5mm}
    \caption{RQ1.1: Accuracy of LLM-generated test cases across HumanEval, MBPP, and APPS datasets using different prompts at test level and task level.}
    \vspace{-5mm}
    \label{fig:rq11}
\end{figure}

The accuracy results of LLM-generated test cases with different prompts are shown in \cref{fig:rq11} and \cref{tab:correctness}. These results reveal that the source code included in prompts significantly affects LLM performance in test case generation at both test level and task level across the HumanEval, MBPP, and APPS datasets. As detailed in \cref{tab:correctness}, prompts incorporating correct code and a task description (\promptsecond) consistently yield the highest accuracy, achieving 80.4\% on HumanEval, 75.4\% on MBPP, and 66.0\% on APPS, averaging 73.9\% across the three datasets at the test level.  In contrast, prompts that incorporate incorrect code with a task description (\promptthird) or correct code without a task description (\promptfour) demonstrate noticeably lower performances, with accuracies of 57.1\% and 64.1\% on HumanEval, 46.3\% and 67.5\% on MBPP, and 37.9\% and 55.3\% on APPS, resulting in average accuracies of 47.1\% and 62.3\%, respectively. Notably, when comparing LLMs provided with incorrect code with task description (\promptthird) against those given correct code and a task description (\promptsecond), there is a 57\% improvement in test accuracy (rising from 47.1\% to 73.9\%). Similarly, compared to the correct code without a task description (\promptfour), \promptsecond yields an 18\% increase in accuracy (from 62.3\% to 73.9\%).

\mybox{\textbf{Answer to RQ1.1}: Incorrect code can significantly impair the ability of LLMs to generate correct tests. Across all three datasets, LLMs achieve approximately 57\% higher accuracy when provided with a task description and correct code (\promptsecond) compared to a task description with incorrect code (\promptthird), improving from 47.1\% to 73.9\%. Among the five prompt types examined, the two most effective are task description alone (\promptfirst) and \promptsecond.}

\subsubsection{RQ1.2: \textbf{Coverage} of LLM-generated test cases}

\begin{figure}
    \centering
    \label{fig:rq1}
    \begin{subfigure}[b]{1\linewidth}
        \centering
        \includegraphics[width=1\linewidth]{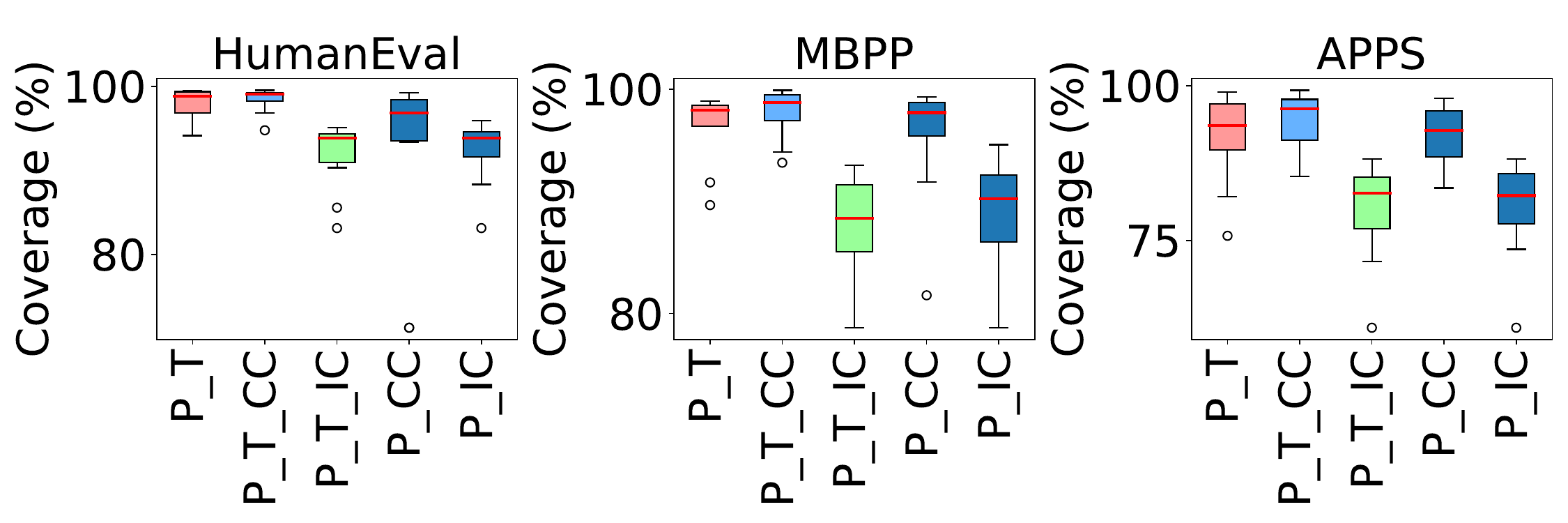}
        \vspace{-5mm}
        \caption{RQ1.2a: Test level.}
        \label{fig:correcttestcoverage}
    \end{subfigure}
    \begin{subfigure}[b]{1\linewidth}
        \centering
        \includegraphics[width=1\linewidth]{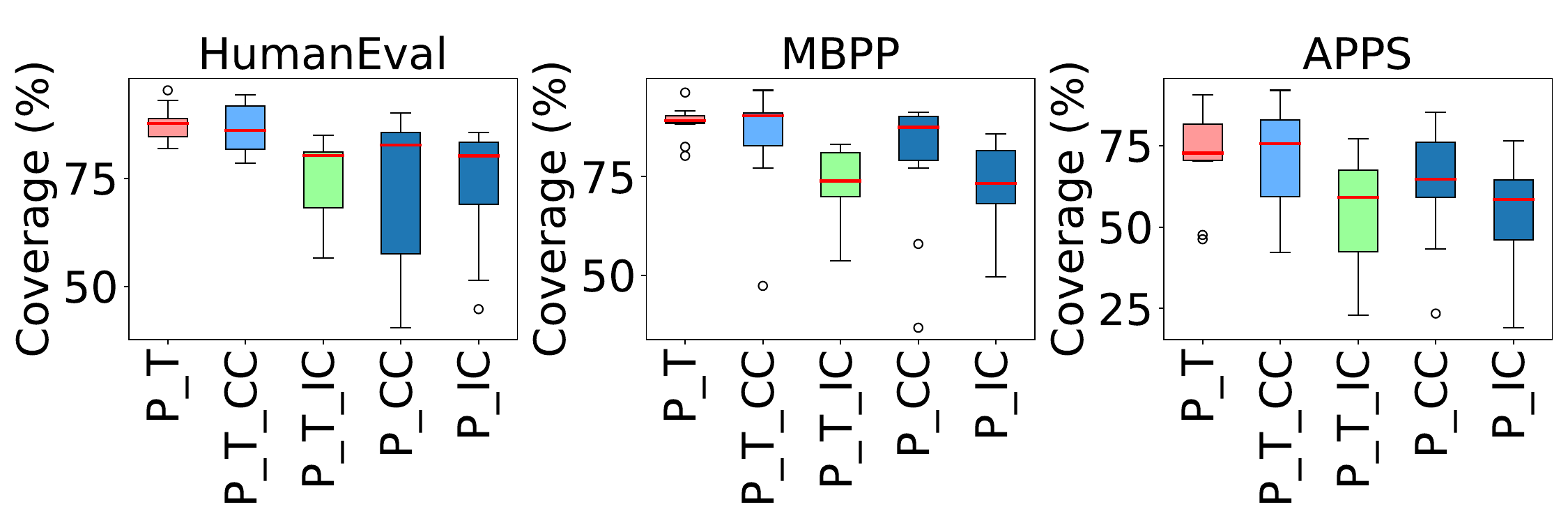}
        \vspace{-5mm}
        \caption{RQ1.2b: Task level.}
        \label{fig:coverage}
    \end{subfigure}
    \vspace{-5mm}
    \caption{RQ1.2: Coverage of LLM-generated test cases.}
    \vspace{-3mm}
    \label{fig:rq12}
\end{figure}

The evaluation results presented in \cref{tab:correcttestcoverage} and \cref{fig:rq12} reveal that the code line coverage achieved by LLM-generated tests is notably influenced by the content of prompts. In particular, similar to test accuracy, both \promptfirst and \promptsecond consistently yield higher coverage compared to the other prompts. As reported in \cref{tab:correcttestcoverage} at the test level, \promptsecond achieves code line coverage rates of 98.4\% on HumanEval, 97.9\% on MBPP, and 94.2\% on APPS, resulting in an average coverage of 96.8\% across the three datasets. In contrast, \promptthird and \promptfour produce lower coverage levels, with HumanEval covering 91.7\% and 92.1\%, MBPP covering 87.7\% and 95.9\%, and APPS covering 80.1\% and 91.9\%, corresponding to average coverages of 86.5\% and 93.3\%, respectively. Notably, when comparing tests generated using incorrect code with a task description (\promptthird) to those using correct code with a task description (\promptsecond), there is a substantial improvement of approximately 12\% in code line coverage (from 86.5\% to 96.8\%). Similarly, the inclusion of a task description alongside correct code in \promptsecond yields an improvement of about 4\% over prompts with correct code but without a task description (\promptfour). These findings underscore that the accuracy of the source code included in the prompts is crucial for enabling LLMs to generate tests with high code coverage.

\mybox{\textbf{Answer to RQ1.2}: Incorrect code also affects the ability of LLMs to generate high-coverage tests. Across all three datasets, LLMs consistently achieve approximately 12\% higher code line coverage when provided with task description and correct code (\promptsecond) compared to task description with incorrect code (\promptthird).}

\begin{table*}
    \centering
    \caption{RQ1.2: Code line coverage of LLM-generated test cases across HumanEval, MBPP, and APPS. 
    Each cell
presents test-level / task-level results.
    Similar to test accuracy, test generations guided by \promptfirst and \promptsecond achieve superior coverage.}
    \resizebox{\textwidth}{!}{
    \begin{tabular}{l|rrrrr|rrrrr|rrrrr}
    \toprule
\multirow{2}{*}{Model}&& & HumanEval &&& & & MBPP & &&& & APPS && \\
 & \promptfirst & \promptsecond & \promptthird&\promptfour&\promptfive & \promptfirst & \promptsecond & \promptthird&\promptfour&\promptfive & \promptfirst & \promptsecond & \promptthird&\promptfour&\promptfive\\
 \midrule
Llama3 & \textbf{95.4} / \textbf{86.0} & 94.8 / 85.0 & 83.2 / 69.6 & 93.6 / 84.0 & 83.2 / 69.6 & 91.7 / \textbf{91.7} & \textbf{94.4} / 87.3 & 78.7 / 73.4 & 91.7 / 82.6 & 78.7 / 73.4 & 82.1 / 77.3 & \textbf{85.4} / \textbf{78.5} & 60.9 / 63.7 & 84.0 / 75.7 & 60.9 / 63.7 \\
CodeLlama& 95.8 / \textbf{83.4} & \textbf{99.1} / 78.6 & 90.3 / 59.4 & 96.8 / 40.4 & 89.9 / 51.4 & 89.7 / 89.5 & \textbf{96.4} / \textbf{96.9} & 84.0 / 66.6 & 81.6 / 81.0 & 83.6 / 59.8 & 75.7 / 83.3 & \textbf{88.8} / \textbf{92.1} & 75.8 / 59.1 & 83.5 / 58.7 & 77.3 / 58.5 \\
DeepSeek & \textbf{99.5} / \textbf{87.4} & 98.2 / 81.6 & 92.0 / 66.7 & 97.2 / 55.9 & 93.2 / 68.4 & \textbf{98.1} / \textbf{88.8} & 98.0 / 47.3 & 86.9 / 58.7 & 96.9 / 36.7 & 89.1 / 66.1 & 89.1 / \textbf{71.5} & \textbf{93.7} / 42.2 & 78.0 / 28.9 & 90.8 / 23.4 & 78.1 / 29.4 \\
StarCoder & \textbf{97.9} / \textbf{95.4} & 96.8 / 94.4 & 85.6 / 56.5 & 93.4 / 46.0 & 88.3 / 44.7 & \textbf{96.7} / \textbf{96.3} & 93.4 / 92.6 & 79.6 / 53.7 & 94.7 / 58.0 & 81.8 / 49.7 & \textbf{90.1} / \textbf{90.7} & 87.7 / 83.7 & 71.6 / 23.0 & 86.2 / 43.3 & 73.6 / 19.1 \\
Codestral & 98.8 / 81.9 & \textbf{99.2} / 80.0 & 95.1 / 80.3 & 98.3 / \textbf{82.9} & 95.9 / 80.7 & 98.6 / 82.6 & 98.4 / 77.2 & 92.0 / 83.2 & \textbf{98.8} / \textbf{87.5} & 93.3 / 73.2 & 93.6 / 46.3 & 93.8 / 50.2 & 88.1 / \textbf{77.2} & \textbf{95.3} / 62.9 & 87.9 / 76.4 \\
GPT3.5 & 98.8 / 87.8 & \textbf{99.0} / \textbf{88.0} & 93.9 / 81.2 & 71.3 / 59.3 & 93.8 / 80.2 & 98.2 / 89.1 & \textbf{98.8} / \textbf{90.5} & 88.5 / 73.5 & 97.9 / 90.3 & 90.7 / 80.6 & 95.7 / 70.4 & \textbf{96.3} / \textbf{75.7} & 82.6 / 43.8 & 93.4 / 64.7 & 82.2 / 45.4 \\
GPT3.5-1106 & 98.9 / 88.6 & \textbf{99.1} / \textbf{91.7} & 94.7 / 83.7 & 71.3 / 61.0 & 93.6 / 82.9 & 97.9 / 90.3 & \textbf{99.0} / \textbf{91.5} & 88.5 / 73.9 & 98.3 / 90.4 & 90.2 / 77.7 & 95.9 / 72.8 & \textbf{97.2} / \textbf{83.2} & 81.7 / 41.0 & 92.8 / 66.4 & 79.2 / 46.8 \\
GPT4-turbo & 99.3 / 88.8 & \textbf{99.5} / \textbf{92.5} & 94.1 / 81.1 & 99.2 / 88.2 & 94.4 / 85.7 & 99.0 / 88.3 & \textbf{99.9} / \textbf{90.7} & 91.6 / 82.1 & 98.8 / 89.2 & 90.1 / 85.9 & 98.9 / 70.8 & \textbf{99.3} / 74.4 & 88.1 / 63.0 & 97.9 / \textbf{76.6} & 86.3 / 60.6 \\
GPT4 & \textbf{99.5} / \textbf{93.1} & 99.3 / 91.8 & 94.0 / 81.0 & 99.2 / 90.2 & 94.7 / 83.8 & 99.0 / 88.9 & \textbf{99.8} / 90.6 & 90.2 / 80.0 & 99.3 / \textbf{91.3} & 91.4 / 82.5 & \textbf{98.8} / 80.7 & \textbf{98.8} / \textbf{82.8} & 83.6 / 57.1 & 97.5 / 77.8 & 85.3 / 57.6 \\
Claude3S & 99.4 / \textbf{83.0} & \textbf{99.5} / 81.9 & 94.5 / 73.8 & 98.6 / 82.8 & 95.4 / 74.9 & 98.6 / 80.3 & \textbf{99.6} / \textbf{81.9} & 93.2 / 79.6 & 98.9 / 77.2 & 94.6 / 70.2 & 98.1 / 47.6 & \textbf{98.4} / 48.1 & 86.4 / \textbf{71.3} & 96.6 / 59.6 & 88.2 / 65.2 \\
Claude3H & 94.1 / \textbf{88.9} & \textbf{98.2} / 86.1 & 91.5 / 85.0 & 94.7 / 87.1 & 94.3 / 85.3 & 96.7 / \textbf{90.7} & \textbf{99.3} / 83.6 & 91.3 / 83.1 & 97.6 / 90.0 & 95.0 / 83.7 & 93.5 / 82.5 & \textbf{96.7} / 68.6 & 83.9 / 76.8 & 92.8 / \textbf{85.3} & 83.2 / 72.7 \\
\midrule
Overall & 97.9 / \textbf{87.7} & \textbf{98.4} / 86.5 & 91.7 / 74.4 & 92.1 / 70.7 & 92.5 / 73.4 & 96.7 / \textbf{88.8} & \textbf{97.9} / 84.6 & 87.7 / 73.4 & 95.9 / 79.5 & 89.0 / 73.0 & 92.0 / \textbf{72.2} & \textbf{94.2} / 70.8 & 80.1 / 55.0 & 91.9 / 63.1 & 80.2 / 54.1 \\
\bottomrule
    \end{tabular}
    }
    \label{tab:correcttestcoverage}
\end{table*}

\begin{table*}
    \centering
    \caption{RQ1.3: Bug detection rate of LLM-generated test cases in our constructed bug set and \promptthird provided bug set. We observe that \promptsecond-guided test cases consistently achieve the highest bug detection effectiveness. 
    }
    \resizebox{\textwidth}{!}{
    \begin{tabular}{l|rrrrr|rrrrr|rrrrr}
    \toprule
\multirow{2}{*}{Model}&& & HumanEval &&& & & MBPP & &&& & APPS && \\
 & \promptfirst & \promptsecond & \promptthird&\promptfour&\promptfive & \promptfirst & \promptsecond & \promptthird&\promptfour&\promptfive & \promptfirst & \promptsecond & \promptthird&\promptfour&\promptfive\\
 \midrule
\multicolumn{10}{l}{Constructed bug set}\\
 \midrule
Llama3 & 41.2 / 28.2 & \textbf{44.7} / \textbf{34.1} & 20.0 / 20.0 & 38.8 / 29.4 & 20.0 / 20.0 & 24.4 / 23.5 & \textbf{29.6} / \textbf{24.4} & 14.6 / 12.2 & 26.3 / 21.1 & 14.6 / 12.2 & 28.5 / 27.9 & \textbf{35.5} / \textbf{33.7} & 8.1 / 32.6 & 30.2 / 27.9 & 8.1 / 32.6 \\
CodeLlama & 44.7 / 18.8 & \textbf{57.6} / 23.5 & 45.9 / \textbf{36.5} & 35.3 / 5.9 & 29.4 / 8.2 & 23.0 / 13.6 & \textbf{35.2} / \textbf{31.0} & 19.7 / 16.9 & 16.4 / 16.4 & 14.1 / 11.3 & 0.0 / 14.0 & \textbf{35.5} / \textbf{33.7} & 12.8 / 16.9 & 27.3 / 14.0 & 17.4 / 19.8 \\
DeepSeek & \textbf{57.6} / \textbf{29.4} & 50.6 / 23.5 & 55.3 / 14.1 & 42.4 / 8.2 & 35.3 / 16.5 & \textbf{39.4} / \textbf{22.5} & 36.2 / 8.5 & 37.1 / 15.0 & 31.5 / 4.7 & 20.7 / 11.3 & 30.2 / \textbf{9.3} & \textbf{58.1} / 6.4 & 48.3 / 8.7 & 36.6 / 2.9 & 20.9 / 3.5 \\
StarCoder & \textbf{51.8} / \textbf{47.1} & 48.2 / \textbf{47.1} & 47.1 / 34.1 & 34.1 / 8.2 & 25.9 / 14.1 & \textbf{32.9} / \textbf{30.5} & 27.7 / 24.9 & 26.3 / 21.1 & 26.8 / 7.5 & 13.6 / 12.2 & \textbf{32.6} / 34.9 & 29.1 / 33.7 & 31.4 / \textbf{36.6} & 25.6 / 4.7 & 15.1 / 20.3 \\
Codestral& 51.8 / 21.2 & \textbf{57.6} / 22.4 & 52.9 / 22.4 & 48.2 / \textbf{27.1} & 36.5 / 21.2 & 35.2 / 17.4 & 36.6 / 13.1 & 34.7 / 17.4 & \textbf{40.4} / \textbf{22.5} & 24.4 / 11.7 & 50.0 / 7.0 & \textbf{52.9} / 9.9 & 45.9 / 6.4 & 47.7 / \textbf{10.5} & 23.8 / 4.7 \\
GPT3.5 & 50.6 / \textbf{25.9} & \textbf{52.9} / 24.7 & 36.5 / 20.0 & 12.9 / 8.2 & 34.1 / 17.6 & 32.4 / 19.7 & \textbf{39.0} / 23.9 & 22.5 / 14.1 & 36.2 / \textbf{25.8} & 22.5 / 16.0 & 47.7 / 16.9 & \textbf{51.2} / \textbf{22.1} & 20.9 / 5.8 & 43.0 / 14.5 & 20.3 / 6.4 \\
GPT3.5-1106 & 49.4 / 28.2 & \textbf{55.3} / \textbf{35.3} & 38.8 / 23.5 & 11.8 / 8.2 & 32.9 / 20.0 & 33.3 / 23.9 & \textbf{40.4} / \textbf{29.1} & 21.6 / 14.1 & 36.2 / 26.3 & 20.7 / 15.0 & 51.7 / 16.3 & \textbf{57.6} / \textbf{29.1} & 20.3 / 5.8 & 41.9 / 16.9 & 19.2 / 7.6 \\
GPT4-turbo & 57.6 / 32.9 & 58.8 / \textbf{37.6} & \textbf{60.0} / 34.1 & 54.1 / 28.2 & 36.5 / 22.4 & 41.8 / 24.4 & \textbf{44.1} / \textbf{29.1} & 39.4 / 19.7 & 39.0 / 23.0 & 19.2 / 10.3 & 63.4 / 17.4 & 65.1 / 22.7 & 63.4 / 16.3 & \textbf{65.7} / \textbf{23.3} & 26.2 / 20.3 \\
GPT4 & 55.3 / \textbf{35.3} & \textbf{57.6} / \textbf{35.3} & 56.5 / 32.9 & 51.8 / 27.1 & 38.8 / 22.4 & 40.8 / 25.8 & \textbf{43.2} / \textbf{28.6} & 40.4 / 24.4 & 39.0 / 27.2 & 24.4 / 18.3 & 63.4 / 25.0 & \textbf{64.0} / 28.5 & 62.8 / 22.7 & 54.1 / 20.3 & 25.0 / \textbf{32.0} \\
Claude3S & 50.6 / 18.8 & \textbf{57.6} / \textbf{23.5} & 56.5 / 18.8 & 48.2 / 21.2 & 35.3 / 10.6 & 36.2 / 17.4 & \textbf{44.6} / \textbf{18.8} & 39.4 / 17.4 & 37.1 / 16.0 & 21.1 / 7.5 & 57.6 / 7.0 & \textbf{64.5} / 9.3 & 62.8 / 8.1 & 52.9 / \textbf{12.2} & 29.1 / 4.7 \\
Claude3H & 17.6 / 4.7 & \textbf{47.1} / 14.1 & 44.7 / \textbf{21.2} & 10.6 / 7.1 & 20.0 / 7.1 & 18.3 / 2.3 & \textbf{35.2} / \textbf{16.0} & 31.0 / 13.1 & 16.9 / 9.9 & 14.1 / 7.5 & 30.8 / 2.3 & \textbf{52.3} / \textbf{14.5} & 39.5 / 8.7 & 13.4 / 8.7 & 10.5 / 11.6 \\
\midrule
Overall & 48.0 / 26.4 & \textbf{53.5} / \textbf{29.2} & 46.7 / 25.2 & 35.3 / 16.3 & 31.3 / 16.4 & 32.5 / 20.1 & \textbf{37.4} / \textbf{22.5} & 29.7 / 16.9 & 31.4 / 18.2 & 19.0 / 12.1 & 41.4 / 16.2 & \textbf{51.4} / \textbf{22.1} & 37.8 / 15.3 & 39.9 / 14.2 & 19.6 / 14.9 \\
\midrule
\multicolumn{10}{l}{Incorrect code from \promptthird}\\
\midrule
Llama3 & 65.9 / 50.6 & \textbf{67.1} / \textbf{52.9} & 28.2 / 31.8 & 63.5 / 49.4 & 28.2 / 31.8 & 47.4 / 47.4 & \textbf{60.6} / \textbf{50.7} & 26.8 / 21.6 & 51.2 / 39.9 & 26.8 / 21.6 & 37.2 / 38.4 & \textbf{47.7} / \textbf{43.6} & 16.9 / 41.3 & 45.9 / 41.9 & 16.9 / 41.3 \\
CodeLlama & 72.9 / 35.3 & \textbf{92.9} / 41.2 & 60.0 / \textbf{50.6} & 76.5 / 9.4 & 48.2 / 12.9 & 45.5 / 30.0 & \textbf{65.7} / \textbf{58.7} & 41.8 / 35.2 & 31.5 / 29.1 & 31.9 / 17.4 & 1.7 / 20.9 & \textbf{47.7} / \textbf{43.0} & 18.0 / 20.9 & 44.2 / 20.3 & 30.8 / 26.2 \\
DeepSeek & \textbf{95.3} / \textbf{48.2} & 85.9 / 37.6 & 88.2 / 29.4 & 81.2 / 15.3 & 57.6 / 21.2 & \textbf{76.5} / \textbf{47.9} & 71.8 / 14.1 & 69.0 / 28.2 & 70.0 / 8.0 & 38.5 / 20.2 & 37.8 / \textbf{11.0} & \textbf{77.3} / 8.1 & 62.8 / 9.3 & 61.6 / 5.2 & 33.1 / 4.7 \\
StarCoder & \textbf{80.0} / \textbf{70.6} & 75.3 / 67.1 & 74.1 / 56.5 & 63.5 / 12.9 & 43.5 / 25.9 & \textbf{65.7} / \textbf{63.4} & 55.9 / 53.5 & 56.3 / 46.9 & 57.3 / 13.1 & 31.0 / 23.9 & 45.3 / \textbf{47.1} & 39.0 / 43.6 & 41.3 / \textbf{47.1} & \textbf{45.9} / 8.1 & 28.5 / 28.5 \\
Codestral & 91.8 / 34.1 & \textbf{97.6} / 36.5 & 92.9 / 36.5 & 88.2 / \textbf{40.0} & 65.9 / 29.4 & 77.5 / 36.6 & \textbf{83.6} / 30.0 & 77.0 / 33.3 & 82.2 / \textbf{46.5} & 48.4 / 20.7 & 70.3 / 11.0 & \textbf{76.2} / 12.8 & 62.2 / 9.9 & 72.1 / \textbf{18.0} & 37.8 / 8.7 \\
GPT3.5 & 89.4 / \textbf{45.9} & \textbf{90.6} / \textbf{45.9} & 60.0 / 32.9 & 24.7 / 16.5 & 61.2 / 30.6 & 76.5 / 49.3 & \textbf{84.0} / \textbf{55.4} & 42.3 / 24.9 & 77.5 / 52.1 & 46.0 / 32.4 & 74.4 / 23.3 & \textbf{75.0} / \textbf{30.2} & 40.1 / 9.3 & 70.3 / 22.7 & 38.4 / 9.3 \\
GPT3.5-1106 & 89.4 / 48.2 & \textbf{92.9} / \textbf{57.6} & 64.7 / 37.6 & 24.7 / 16.5 & 62.4 / 35.3 & 72.8 / 48.8 & \textbf{81.7} / \textbf{57.3} & 40.8 / 24.4 & 78.4 / 52.6 & 44.1 / 28.6 & 75.0 / 25.6 & \textbf{77.9} / \textbf{41.3} & 37.2 / 8.7 & 66.3 / 23.8 & 34.3 / 10.5 \\
GPT4-turbo & 95.3 / 47.1 & \textbf{97.6} / \textbf{57.6} & \textbf{97.6} / 50.6 & 90.6 / 47.1 & 63.5 / 32.9 & 83.1 / 46.5 & \textbf{89.7} / \textbf{55.4} & 76.5 / 37.1 & 81.7 / 45.1 & 41.8 / 21.6 & 87.8 / 24.4 & \textbf{90.7} / 30.8 & 87.8 / 22.7 & \textbf{90.7} / \textbf{32.0} & 43.6 / 26.7 \\
GPT4 & \textbf{95.3} / \textbf{57.6} & \textbf{95.3} / 55.3 & 92.9 / 52.9 & 91.8 / 50.6 & 64.7 / 37.6 & 79.8 / 47.9 & \textbf{85.9} / 53.1 & 79.3 / 47.4 & 85.0 / \textbf{56.3} & 49.3 / 36.2 & 87.8 / 36.0 & \textbf{89.5} / 36.0 & \textbf{89.5} / 31.4 & 82.6 / 33.1 & 44.2 / \textbf{43.6} \\
Claude3S & 92.9 / 34.1 & \textbf{97.6} / \textbf{38.8} & \textbf{97.6} / 28.2 & 85.9 / 36.5 & 62.4 / 21.2 & 79.8 / 33.8 & \textbf{85.9} / \textbf{34.7} & 81.7 / 31.0 & 81.2 / 28.6 & 46.0 / 13.1 & 84.9 / 11.6 & \textbf{89.5} / 12.8 & 87.8 / 12.8 & 77.3 / \textbf{17.4} & 48.8 / 8.7 \\
Claude3H & 25.9 / 8.2 & \textbf{68.2} / 20.0 & \textbf{68.2} / \textbf{29.4} & 18.8 / 10.6 & 28.2 / 9.4 & 32.9 / 4.7 & \textbf{70.0} / 26.3 & 68.1 / \textbf{27.2} & 31.0 / 15.0 & 23.5 / 13.1 & 41.9 / 6.4 & \textbf{75.6} / \textbf{22.1} & 59.9 / 14.5 & 21.5 / 12.2 & 19.2 / 14.0 \\
\midrule
Overall & 81.3 / 43.6 & \textbf{87.4} / \textbf{46.4} & 75.0 / 39.7 & 64.5 / 27.7 & 53.3 / 26.2 & 67.1 / 41.5 & \textbf{75.9} / \textbf{44.5} & 60.0 / 32.5 & 66.1 / 35.1 & 38.8 / 22.6 & 58.6 / 23.3 & \textbf{71.5} / \textbf{29.5} & 54.9 / 20.7 & 61.7 / 21.4 & 34.1 / 20.2 \\
\bottomrule
    \end{tabular}
    }
    \label{tab:correcttestbugdetection}
\end{table*}

\subsubsection{RQ1.3: \textbf{Bug detection effectiveness} of LLM-generated test cases }

To evaluate the effectiveness of LLM-generated test cases in identifying errors within faulty implementations, we assessed their bug detection capabilities using both our constructed bug set and the buggy code provided by \promptthird. As shown in \cref{tab:correcttestbugdetection}, \promptsecond consistently outperforms other prompts across all three datasets. Specifically, for the constructed bug set, LLMs using \promptsecond achieved detection rates of 53.5\% on HumanEval, 37.4\% on MBPP, and 51.4\% on APPS, yielding an average detection rate of 47.4\%. In comparison, test cases generated with \promptthird recorded detection rates of 46.7\% on HumanEval, 29.7\% on MBPP, and 37.8\% on APPS for our constructed bug set (averaging 38.1\%). These results indicate that \promptsecond outperforms \promptthird by roughly 24\% (from 38.1\% to 47.4\%) in bug detection effectiveness for our constructed bug sets. Moreover, when comparing \promptsecond with \promptfour (provides only correct code implementations without task descriptions), \promptfour achieved significantly lower detection rates of 35.3\% on HumanEval, 31.4\% on MBPP, and 39.9\% on APPS for the constructed bug set (averaging 35.5\%), corresponding to a 34\% (from 35.5\% to 47.4\%) performance gap relative to \promptsecond. Overall, these findings suggest that LLM-generated test cases based on \promptsecond yield the highest bug detection effectiveness across all datasets.

\mybox{\textbf{Answer to RQ1.3}: LLM-generated test cases based on task descriptions with correct code (\promptsecond) achieve the highest bug detection across all datasets. On average, this approach achieves approximately 24\% higher bug detection rates compared to using task descriptions with incorrect code (\promptthird).}

\subsection{RQ2: How does the source of the code influence the LLMs in test generation?}

To explore whether LLMs are more easily misled by the code they generate themselves (Own) compared to directly using source code produced elsewhere (Others), for each LLM, we compare the accuracy of test cases generated with 1) \promptsecond with correct code produced elsewhere; 2) \promptsecond with correct code generated by itself; 3) \promptthird with incorrect code produced elsewhere; 4) \promptthird with incorrect code generated by its own. Then we report the evaluation results by calculating the \textbf{diff\_absolute} between the accuracy of \promptsecond ~- the accuracy of \promptthird, and \textbf{diff\_relative}, i.e., diff\_absolute/accuracy of \promptsecond. The comparison is based on identical coding tasks. \cref{tab:rq3} summarizes the results for both \textbf{diff\_absolute} and \textbf{diff\_relative} across three datasets. The outcomes indicate that when using externally sourced code, both diff\_absolute and diff\_relative are markedly higher than when self-generated code is employed. For example, at the test level, diff\_absolute for \textit{Others} is 10.5\% on HumanEval, 23.2\% on MBPP, and 17.5\% on APPS, which averages to 17.1\% across datasets. In contrast, for \textit{Own}, diff\_absolute is 7.7\% on HumanEval, 9.6\% on MBPP, and 8.2\% on APPS, with an overall average of 8.5\%. Thus, on average, employing self-generated code reduces the diff\_absolute by approximately 50\% relative to using externally sourced code.

\mybox{\textbf{Answer to RQ2}: LLMs demonstrate a reduced susceptibility to errors when working with self-generated code. Across all datasets, the use of self-generated code lowers test accuracy differences by roughly 50\% (from 17.1\% to 8.5\%).}

\begin{table*}
    \centering
\caption{RQ2: Test accuracy differences for different sources of code. 
Comparing external (Column ``Others'') vs. self-generated (Column ``Own'') sources. We compute diff\_absolute (accuracy difference between \promptsecond and \promptthird)  diff\_relative (diff\_absolute / \promptsecond). The results suggest that suggesting LLMs are less misled by self-generated code.
}
\vspace{-3mm}
    \resizebox{\textwidth}{!}{
    \begin{tabular}{l|rrrr|rrrr|rrrr}
    \toprule
\multirow{3}{*}{Model}&\multicolumn{4}{|c}{HumanEval}&\multicolumn{4}{|c}{MBPP}&\multicolumn{4}{|c}{APPS} \\
& \multicolumn{2}{|c}{diff\_absolute}& \multicolumn{2}{c}{diff\_relative}& \multicolumn{2}{|c}{diff\_absolute}& \multicolumn{2}{c}{diff\_relative}& \multicolumn{2}{|c}{diff\_absolute}& \multicolumn{2}{c}{diff\_relative}\\
&Others&Own&Others&Own&Others&Own&Others&Own&Others&Own&Others&Own\\
 \midrule
    Llama3 & \textbf{76.2} / \textbf{83.3} & 31.4 / 77.4 & \textbf{76.2} / \textbf{83.3} & 31.4 / 77.4 & \textbf{37.5} / \textbf{46.5} & 11.8 / 11.2 & \textbf{62.6} / \textbf{77.6} & 18.9 / 27.9 & \textbf{-19.1} / -3.5 & -51.7 / \textbf{-15.1} & \textbf{0.0} / \textbf{0.0} & 0.0 / \textbf{0.0}\\
    CodeLlama  & \textbf{-24.9} / \textbf{17.5} & -42.4 / 16.3 & \textbf{-51.4} / \textbf{29.2} & -109.8 / \textbf{40.6} & \textbf{36.2} / \textbf{21.0} & -27.8 / -15.1 & \textbf{45.7} / \textbf{33.7} & -144.6 / \textbf{0.0} & \textbf{28.9} / \textbf{19.3} & 19.6 / 13.9 & \textbf{31.8} / \textbf{36.7} & 40.1 / \textbf{44.1}\\
    DeepSeek& \textbf{9.6} / \textbf{28.0} & -2.9 / -1.5 & \textbf{12.4} / \textbf{70.4} & -4.2 / -9.1 & 13.0 / -17.2 & \textbf{15.7} / \textbf{-3.1} & 19.2 / -103.9 & \textbf{21.8} / \textbf{-23.3} & 0.2 / \textbf{0.4} & \textbf{13.1} / -1.8 & 0.4 / \textbf{3.2} & \textbf{23.4} / -29.0\\
    StarCoder & 18.2 / 41.0 & \textbf{41.1} / \textbf{48.2} & 18.2 / 41.0 & \textbf{41.1} / \textbf{48.2} & \textbf{-12.5} / \textbf{17.2} & -28.4 / -2.0 & \textbf{-30.9} / \textbf{24.0} & -93.6 / -3.2 & -8.5 / -16.4 & \textbf{21.0} / \textbf{14.0} & -16.9 / -49.1 & \textbf{24.0} / \textbf{21.0}\\
    Codestral  & 2.4 / -4.4 & \textbf{5.9} / \textbf{-5.9} & 2.9 / -12.0 & \textbf{7.9} / \textbf{-20.0} & 18.4 / \textbf{13.3} & \textbf{20.5} / 17.2 & 24.7 / \textbf{36.0} & \textbf{29.2} / \textbf{63.9} & \textbf{22.5} / \textbf{14.0} & 18.0 / 9.0 & \textbf{35.7} / \textbf{78.2} & 31.0 / \textbf{77.6}\\
    GPT3.5 & \textbf{26.4} / \textbf{28.6} & 16.8 / 30.1 & \textbf{31.9} / \textbf{53.4} & 20.8 / \textbf{54.6} & \textbf{47.2} / \textbf{47.8} & 23.5 / 27.8 & \textbf{55.7} / \textbf{72.6} & 28.7 / 43.5 & \textbf{47.6} / \textbf{33.5} & 13.3 / 27.9 & \textbf{75.0} / \textbf{90.4} & 21.3 / 63.5\\
    GPT3.5-turbo & 10.8 / 13.8 & \textbf{21.1} / \textbf{24.9} & 13.0 / 23.7 & \textbf{24.8} / \textbf{42.7} & \textbf{49.8} / \textbf{55.6} & 28.0 / 36.3 & \textbf{58.0} / \textbf{78.8} & 33.2 / 53.4 & \textbf{53.4} / \textbf{39.2} & 17.5 / 15.8 & \textbf{75.3} / \textbf{87.2} & 27.4 / 45.1\\
    GPT4-turbo & \textbf{5.6} / \textbf{8.0} & -2.0 / 5.6 & \textbf{6.2} / \textbf{13.8} & -2.2 / 10.0 & \textbf{16.8} / \textbf{36.6} & 14.4 / 30.4 & \textbf{19.4} / \textbf{59.8} & 16.5 / 50.7 & \textbf{19.0} / \textbf{23.1} & 12.0 / 15.7 & \textbf{25.5} / \textbf{72.3} & 16.1 / 47.0\\
    GPT-4  & -6.6 / -15.0 & \textbf{16.1} / \textbf{11.0} & -7.4 / -26.6 & \textbf{18.3} / \textbf{20.4} & \textbf{13.1} / \textbf{22.7} & 12.7 / 22.1 & \textbf{14.9} / \textbf{35.5} & 15.6 / 34.8 & \textbf{23.3} / \textbf{24.7} & 11.8 / 10.2 & \textbf{30.6} / \textbf{58.9} & 16.3 / 24.8\\
    Claude3S & -0.6 / \textbf{11.5} & \textbf{-0.5} / -1.3 & -0.7 / \textbf{28.4} & \textbf{-0.7} / -4.7 & \textbf{15.5} / \textbf{12.9} & 12.2 / 16.8 & \textbf{19.1} / \textbf{32.4} & 15.9 / 47.5 & \textbf{21.5} / \textbf{21.4} & 4.7 / 15.2 & \textbf{28.7} / \textbf{64.2} & 7.8 / 63.8\\
    Claude3H & -1.9 / \textbf{9.4} & \textbf{0.2} / -7.3 & -2.2 / \textbf{18.3} & \textbf{0.3} / -20.8 & 20.5 / \textbf{22.6} & \textbf{23.1} / 27.9 & 27.6 / \textbf{55.5} & \textbf{29.9} / 67.1 & 3.4 / -3.9 & \textbf{10.7} / \textbf{15.0} & 4.8 / -11.7 & \textbf{15.5} / \textbf{46.9}\\
    \midrule
    Overall & \textbf{10.5} / \textbf{20.2} & 7.7 / 18.0 & \textbf{12.5} / \textbf{29.3} & 9.6 / 21.8 & \textbf{23.2} / \textbf{25.4} & 9.6 / 15.4 & \textbf{31.1} / \textbf{36.5} & 14.2 / 32.9 & \textbf{17.5} / \textbf{13.8} & 8.2 / 10.9 & \textbf{27.8} / \textbf{39.1} & 13.8  / 36.8 \\
\bottomrule
    \end{tabular}
    }
    \label{tab:rq3}
\end{table*}

\subsection{RQ3: To what extent are LLMs misguided by the incorrect code in test generation?}

To answer this RQ, we compared the pass rates of test cases produced by \promptfour and \promptfive when evaluated on the incorrect implementations provided by \promptfive. 
In other words, we measure the ratio of tests that use the behaviours of incorrect code as test oracles.  
The results reported in \cref{tab:bias} indicate that LLMs tend to align with the code presented in the prompt, thereby generating test cases that erroneously pass the incorrect implementations. For example, test cases generated with \promptfour yield pass rates of 30.7\%, 25.3\%, and 22.8\% on the incorrect code in the HumanEval, MBPP, and APPS datasets, respectively, which corresponds to an average pass rate of 26.3\%. In contrast, tests generated with \promptfive attain higher pass rates on incorrect code with 41.5\%, 30.7\%, and 32.3\% pass@1 on the same datasets, averaging 34.8\% pass@1. Consequently, when comparing the LLMs provided with correct code (\promptfour) against those given incorrect code (\promptfive), there is an observed 24\% reduction in the average pass rate (from 34.8\% to 26.3\%).

\mybox{\textbf{Answer to RQ3}: Erroneous source code in prompts misguides LLMs, leading them to generate a greater proportion of test cases that inappropriately pass faulty implementations. Across all three datasets, test cases generated with correct code (\promptfour) achieve an average pass rate that is 24\% lower than that of those generated with incorrect code (\promptfive).}

\begin{table}
    \centering
    \caption{RQ3: Pass rate of the LLM-generated test cases on the incorrect code provided by \promptfive. 
    Tests generated by prompts containing incorrect code are observed to have higher pass rates on the incorrect code at both the test and task levels.
    }
    \resizebox{1\linewidth}{!}{
    \begin{tabular}{l|rrrr|rrrr|rrrr}
    \toprule
\multirow{2}{*}{Model}&\multicolumn{4}{|c}{HumanEval}&\multicolumn{4}{|c}{MBPP}&\multicolumn{4}{|c}{APPS} \\
&\multicolumn{2}{|c}{Test Level}&\multicolumn{2}{c}{Task Level}&\multicolumn{2}{|c}{Test Level}&\multicolumn{2}{c}{Task Level}&\multicolumn{2}{|c}{Test Level}&\multicolumn{2}{c}{Task Level}\\
\midrule
 & \promptfour& \promptfive & \promptfour& \promptfive& \promptfour& \promptfive& \promptfour& \promptfive & \promptfour& \promptfive& \promptfour& \promptfive\\
\midrule
Llama3 & 9.6 & \textbf{19.6} & 15.3 & \textbf{36.5} & 17.1 & \textbf{20.4} & 21.6 & \textbf{39.4} & 3.5 & \textbf{19.1} & 15.7 & \textbf{61.0} \\
CodeLlama & 27.8 & \textbf{30.7} & 5.9 & \textbf{18.8} & 16.2 & \textbf{20.9} & \textbf{58.2} & 27.2 & 12.9 & \textbf{19.7} & 29.6 & \textbf{43.6} \\
DeeoSeek & 30.9 & \textbf{39.1} & 10.6 & \textbf{23.5} & 21.9 & \textbf{26.6} & 7.5 & \textbf{23.9} & 18.7 & \textbf{21.6} & 5.8 & \textbf{10.5} \\
StarCoder & 22.9 & \textbf{23.8} & 10.6 & \textbf{23.5} & 21.9 & \textbf{27.6} & 10.3 & \textbf{23.0} & 14.4 & \textbf{23.6} & 5.2 & \textbf{28.5} \\
Codestral & 45.5 & \textbf{46.6} & 20.0 & \textbf{22.4} & 28.6 & \textbf{29.8} & \textbf{18.3} & 17.8 & 23.2 & \textbf{30.9} & 3.5 & \textbf{18.6} \\
GPT3.5 & 8.4 & \textbf{41.2} & 5.9 & \textbf{24.7} & 27.5 & \textbf{31.6} & 20.2 & \textbf{23.5} & 21.8 & \textbf{28.3} & 7.6 & \textbf{14.5} \\
GPT3.5-1106 & 10.0 & \textbf{39.7} & 8.2 & \textbf{27.1} & 27.1 & \textbf{31.6} & 19.2 & \textbf{24.4} & \textbf{25.8} & 24.5 & 9.9 & \textbf{13.9} \\
GPT4-turbo & 44.3 & \textbf{58.9} & 27.1 & \textbf{45.9} & 31.5 & \textbf{42.8} & 23.5 & \textbf{43.2} & 29.6 & \textbf{55.5} & 7.6 & \textbf{39.5} \\
GPT4 & 43.6 & \textbf{55.9} & 23.5 & \textbf{38.8} & 28.1 & \textbf{40.1} & 22.1 & \textbf{38.0} & 26.1 & \textbf{49.7} & 12.8 & \textbf{51.7} \\
Claude3S & 39.6 & \textbf{46.2} & 20.0 & \textbf{22.4} & 24.2 & \textbf{31.5} & 11.3 & \textbf{17.4} & 21.5 & \textbf{35.8} & 4.1 & \textbf{25.0} \\
Claude3H & 54.7 & \textbf{54.8} & \textbf{42.4} & 35.3 & 34.4 & \textbf{35.1} & \textbf{31.5} & 29.1 & \textbf{53.4} & 46.5 & \textbf{42.4} & 40.1 \\
\midrule
Overall & 30.7 & 41.5 & 17.2 & 29.0 & 25.3 & 30.7 & 22.1 & 27.9 & 22.8 & 32.3 & 13.1 & 31.6 \\
\bottomrule
    \end{tabular}
    }
    \label{tab:bias}
\end{table}

\begin{table}
\centering
\caption{RQ4: Evaluation results of three metrics at test level across three datasets in GPT-4 generated test cases. HE denotes HumanEval.}
\label{tab:deviation_impact}
\resizebox{1\linewidth}{!}{%
\begin{tabular}{l|rrr|rrr|rrr|rrr}
\midrule
\multirow{2}{*}{Deviation} & \multicolumn{3}{|c}{CodeBLEU}& \multicolumn{3}{|c}{Accuracy (\%)} & \multicolumn{3}{|c}{Coverage (\%)} & \multicolumn{3}{|c}{Bug Detection (\%)}  \\
 & HE & MBPP & APPS & HE & MBPP & APPS & HE & MBPP & APPS & HE & MBPP & APPS \\
\toprule
Min & 0.38 & 0.39 & 0.62 & 75.7 & 64.0 & 52.8 & 92.2 & 88.4 & 82.3 & 58.8 & 37.6 & 59.9 \\
\textit{Main} & 0.58 & 0.65 & 0.67 & 76.2 & 64.9 & 53.6 & 94.0 & 90.2 & 83.6 & 56.5 & 40.4  & 62.8\\
Max  & 0.75 & 0.85 & 0.95 & \textbf{78.1} & \textbf{66.5} & \textbf{56.6} & \textbf{95.2} & \textbf{92.7} & \textbf{85.9} & \textbf{61.2} & \textbf{41.8} & \textbf{69.8}\\
\bottomrule
\end{tabular}
}
\end{table}

\subsection{RQ4: How does code incorrectness degree impact test case generation?}

To explore the influence of code incorrectness degree, we measured the deviation between LLM-generated incorrect code, and then analyzed the effectiveness of LLM-generated test cases under different levels of code deviation.  \cref{tab:deviation_impact} presents the evaluation results of LLM-generated test cases under \promptthird with three degrees of deviation, measured by CodeBLEU between correct and incorrect code, where \textit{Main} mean results reported in \cref{sec:experiment}. Our analysis indicates that, in most cases, higher CodeBLEU scores (i.e., greater similarity between the correct and incorrect code) correlate with improvements in the quality of the generated test cases. For example, when using incorrect code with the largest CodeBLEU, the accuracy of LLM-generated test cases has increased from 75.7\% to 78.1\%. 

\mybox{\textbf{Answer to RQ4}: 
As the code more closely resembles a correct implementation, the generated test cases are improved across all metrics.
For example, for HumanEval, raising the CodeBLEU score from 0.38 to 0.75 leads to an increase in test accuracy (from 75.7\% to 78.1\%), coverage (from 92.2\% to 95.2\%), and bug detection rate (from 58.8\% to 61.2\%).}

\begin{table}
    \centering
    \caption{RQ5: Effectiveness of test cases with BugsInPy and SWE-Bench datasets. 
Averaged across all LLMs, \promptfour achieves 19.3\% test accuracy, 13.6\% coverage, and 13.5\% bug detection, outperforming \promptfive's 13.7\%, 7.2\%, and 7.1\%.}
\resizebox{1\linewidth}{!}{
    \begin{tabular}{l|rrrr|rrrr|rrrr}
    \toprule
\multirow{3}{*}{Model}&\multicolumn{4}{|c}{Accuracy}&\multicolumn{4}{|c}{Coverage}&\multicolumn{4}{|c}{Bug Detection} \\
&\multicolumn{2}{|c}{Test Level}&\multicolumn{2}{c}{Task Level}&\multicolumn{2}{|c}{Test Level}&\multicolumn{2}{c}{Task Level}&\multicolumn{2}{|c}{Test Level}&\multicolumn{2}{c}{Task Level}\\
\midrule
 & \promptfour& \promptfive & \promptfour& \promptfive& \promptfour& \promptfive& \promptfour& \promptfive& \promptfour& \promptfive& \promptfour& \promptfive\\
\midrule

Llama3 & 5.9 & \textbf{10.0} & \textbf{4.9} & 4.9 & \textbf{21.2} & 21.0 & \textbf{0.5} & 0.5 & \textbf{31.7} & 31.7 & \textbf{2.4} & 2.4 \\
CodeLlama & \textbf{18.6} & 12.3 & \textbf{7.3} & 4.9 & \textbf{21.3} & 21.1 & \textbf{1.0} & 0.6 & \textbf{31.7} & 29.3 & \textbf{2.4} & 0.0 \\
DeepSeek & 9.8 & \textbf{16.3} & \textbf{9.8} & 9.8 & \textbf{21.3} & 21.1 & \textbf{9.3} & 9.3 & \textbf{29.3} & 29.3 & \textbf{4.9} & 4.9 \\
StarCoder & 1.0 & \textbf{1.2} & 2.4 & \textbf{4.9} & \textbf{21.0} & 21.0 & 0.4 & \textbf{0.5} & \textbf{31.7} & 31.7 & \textbf{2.4} & 2.4 \\
Codestral & \textbf{42.1} & 40.7 & \textbf{24.4} & 22.0 & \textbf{21.4} & 21.3 & \textbf{14.6} & 14.6 & \textbf{36.6} & 31.7 & \textbf{12.2} & 7.3 \\
GPT3.5 & 15.9 & \textbf{16.3} & \textbf{12.2} & 12.2 & \textbf{25.2} & 25.1 & \textbf{5.1} & 5.1 & \textbf{24.4} & 24.4 & \textbf{2.4} & 2.4 \\
GPT3.5-1106 & \textbf{9.6} & 7.3 & \textbf{7.3} & 4.9 & \textbf{25.2} & 24.9 & \textbf{1.8} & 1.4 & \textbf{24.4} & 24.4 & \textbf{2.4} & 2.4 \\
GPT4-turbo & \textbf{51.5} & 27.3 & \textbf{39.0} & 22.0 & \textbf{25.8} & 25.2 & \textbf{34.6} & 16.0 & \textbf{41.5} & 31.7 & \textbf{29.3} & 4.9 \\
GPT4 & \textbf{18.3} & 12.6 & \textbf{24.4} & 14.6 & \textbf{25.6} & 25.2 & \textbf{16.7} & 11.1 & \textbf{31.7} & 29.3 & \textbf{12.2} & 9.8 \\
Claude3S & \textbf{29.0} & 23.4 & \textbf{31.7} & 29.3 & 25.3 & \textbf{25.4} & \textbf{23.3} & 16.8 & \textbf{43.9} & 39.0 & \textbf{29.3} & 24.4 \\
Claude3S & \textbf{60.6} & 24.7 & \textbf{48.8} & 22.0 & \textbf{25.2} & 25.0 & \textbf{42.8} & 3.8 & \textbf{61.0} & 39.0 & \textbf{48.8} & 17.1 \\
\midrule
Overall & \textbf{23.9} & 17.5 & \textbf{19.3} & 13.7 & \textbf{23.5} & 23.3 & \textbf{13.6} & 7.2 & \textbf{35.3} & 31.0 & \textbf{13.5} & 7.1 \\

\bottomrule
    \end{tabular}
    }
    \label{tab:real}
\end{table}

\subsection{RQ5: Do our observations hold for real-world code?}\label{sec:real-world}

To determine whether our initial findings extend to real-world scenarios, we conducted experiments using tasks collected from two benchmarks: 10 tasks from BugsInPy \cite{widyasari2020bugsinpy} and 31 tasks from SWE-bench \cite{jimenez2023swe}. Because the collected functions could only be processed using \promptfour and \promptfive, we compare the evaluation results of these two strategies. As shown in \cref{tab:real}, our results in real-world settings are consistent with previous observations. In most experiments, test cases generated using \promptfour exhibited higher effectiveness compared to those produced with \promptfive. Specifically, when using \promptfour, LLMs achieved an average task-level accuracy of 19.3\%, code line coverage of 13.6\%, and bug detection rate of 13.5\%. In contrast, \promptfive yielded an average accuracy of 13.7\%, code line coverage of 7.2\%, and bug detection rate of 7.1\%. Notably, the use of incorrect code with \promptfive significantly compromised the performance of the generated test cases, with the bug detection rate approximately 47\% lower than that obtained with \promptfour.

\mybox{\textbf{Answer to RQ5}: 
Our observations hold for real-world code. Across the evaluated tasks, LLMs achieve a 47\% lower bug detection rate when presented with incorrect code (\promptfive) compared to correct code (\promptfour).}

\begin{figure}
    \centering

        \includegraphics[width=\linewidth]{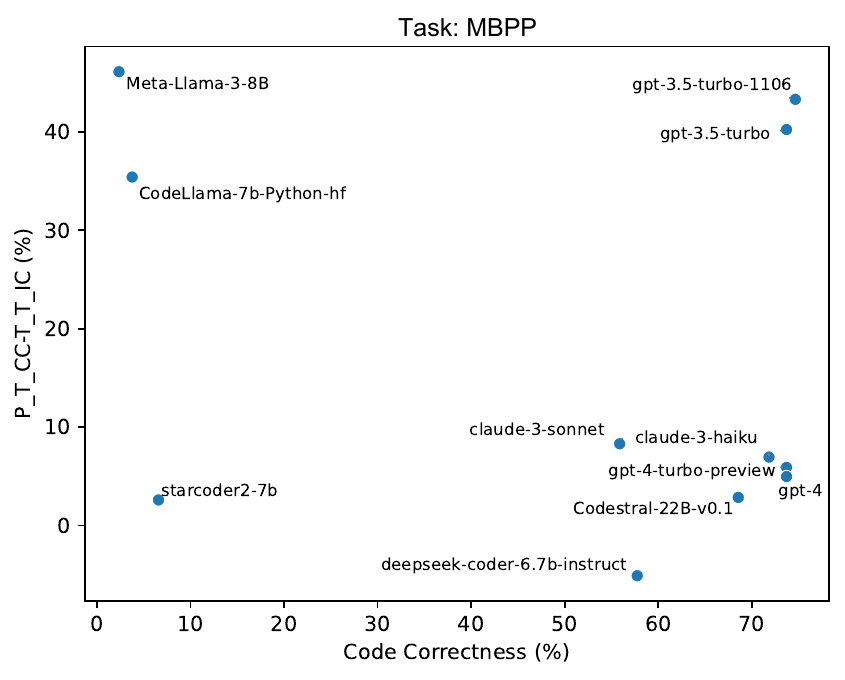}
        \vspace{-5mm}
    \caption{Correlation between the code generation capability of LLMs and their ease of being misled during test generation in the MBPP dataset.}
    \label{fig:correlation}
\end{figure}

\section{Discussion and Extended Analysis}
This section presents an extended analysis of our findings. Due to space constraints, we provide one representative model result for each subsection, offering a focused examination of our research outcomes.

\subsection{Correlation between LLM code generation capability and their ease of being misled}

\cref{fig:correlation} presents the correlation between the code generation capability of LLMs (in terms of pass@1) on the MBPP dataset (zero-shot results) and the difference in performance between \promptsecond and \promptthird\footnote{Additional dataset results and analyses are available in our anonymous GitHub repository (link provided at the end of \cref{sec:conclusion}).}. We observe no strong correlation between an LLM's code generation capability and its susceptibility to being misled during test generation. As shown in \cref{fig:correlation}, even as the code correctness of LLM-generated code increases from 0\% to over 80\%, the difference between \promptfour and \promptfive (\promptfour - \promptfive) randomly distributes from 0\% to 10\%. This finding suggests that an LLM's proficiency in generating correct code does not necessarily translate to a higher resistance to misleading test cases.

\subsection{Incorrect test inputs v.s. test oracles}

We categorize incorrect tests generated by LLMs into two types: incorrect test inputs and incorrect test oracles. Their distribution is presented in \cref{table:error_distribution}. We observe that the majority of incorrect tests are due to incorrect test oracles.
For \promptfive, there is a much higher percentage of flawed test inputs (48.6\%), indicating that supplying incorrect code without any task description shifts GPT-4's mistakes toward the test inputs.

\begin{table}
\centering
\caption{Distribution (\%) of incorrect test inputs and oracles for incorrect test cases generated by GPT-4 for HumanEval.}
\resizebox{0.9\linewidth}{!}{
\begin{tabular}{l|rrrrr}
\toprule
Goal & \promptfirst&\promptsecond&\promptthird&\promptfour&\promptfive \\
\midrule
Test Input & 0.7 & 3.2 & 2.1 & 5.2 & 48.6 \\
\midrule
Test Oracle & 99.3 & 96.8 & 97.9 & 94.8 & 51.4 \\
\bottomrule
\end{tabular}}
\label{table:error_distribution}
\end{table}

\subsection{Results for other programming languages}\label{sec:multilanguage}
To investigate whether our observations hold for non-python languages, we have conducted an empirical study using the HumanEval-X (C++) and HumanEval-X (Java) subsets, which consist of 125 and 100 tasks, respectively. As shown in \cref{tab:task_level_accuracy}, we can observe that the overall trends have been similar to our original results, with \promptsecond and \promptfirst achieving the highest accuracy compared to other prompts. For instance, LLM-generated test cases achieve 33.6\% and 81.0\% task-level accuracy with \promptsecond for C++ and Java, respectively, compared to only 23.2\% and 71.0\% with \promptthird.

\begin{table}
\caption{Accuracy and coverage of the GPT-3.5-turbo-1106-generated code for different prompts in the HumanEval-X (C++ and Java). We have used \textbf{Gcov} and \textbf{JaCoCo} for the code line coverage of LLM-generated C++ and Java code.}
\centering
\resizebox{0.9\linewidth}{!}{
\begin{tabular}{l|rrrrr}
\toprule
Model & P\_T & P\_T\_CC & P\_T\_IC & P\_CC & P\_IC \\
\midrule
\multicolumn{6}{l}{Task level Accuracy} \\
\midrule
C++ & 29.6\% & 33.6\% & 23.2\% & 27.2\% & 18.4\% \\
Java & 76.0\% & 81.0\% & 71.0\% & 72.0\% & 63.0\% \\
\midrule
\multicolumn{6}{l}{Task level Coverage} \\
\midrule
C++ & 47.5\% & 59.3\% & 42.7\% & 45.4\% & 37.3\% \\
Java & 85.1\% & 87.6\% & 81.1\% & 85.1\% & 77.8\% \\
\bottomrule
\end{tabular}}
\label{tab:task_level_accuracy}
\end{table}

\begin{figure}
    \centering
    \includegraphics[width=1\linewidth]{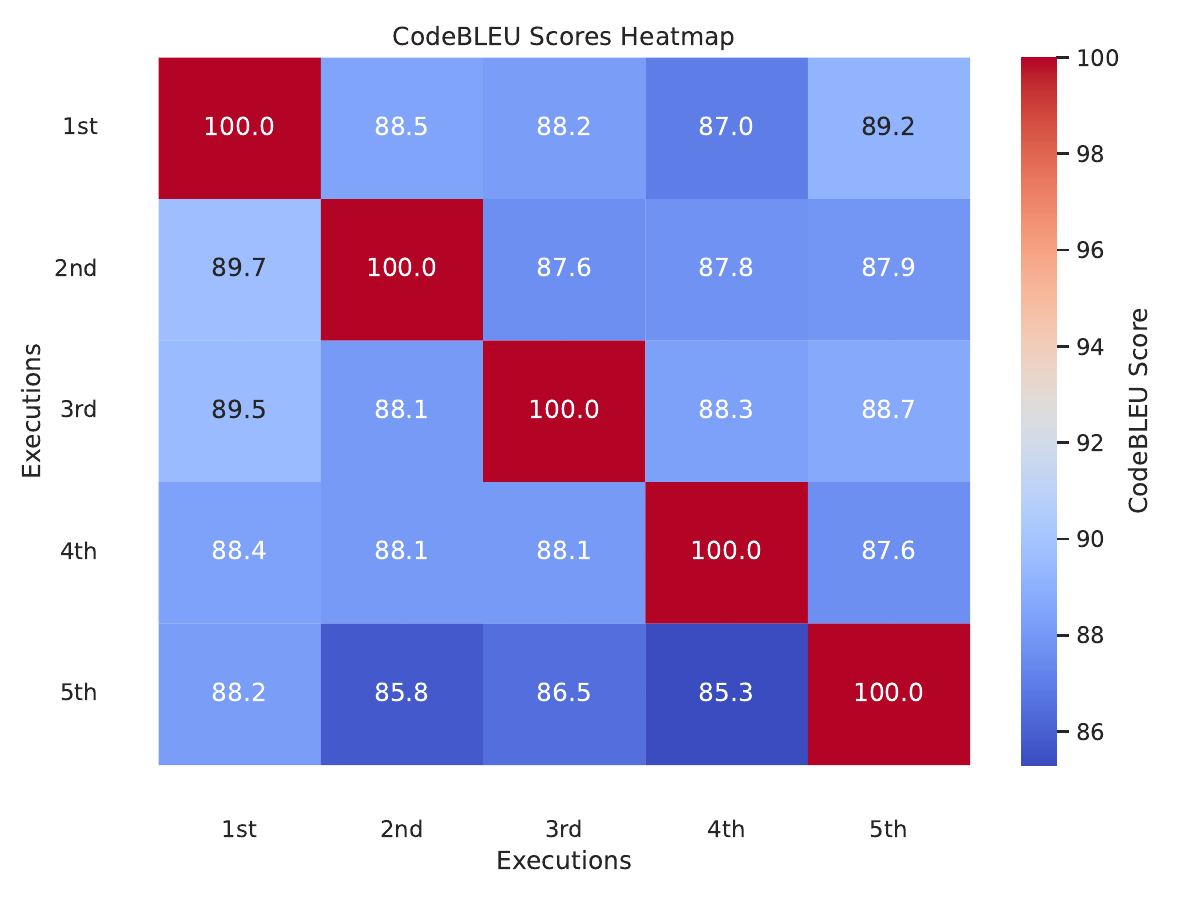}
    \vspace{-3mm}
   \caption{CodeBLEU scores of GPT-3.5-turbo generated test cases across five executions.}
   \vspace{-3mm}
    \label{fig:codebleu}
\end{figure}

\subsection{Randomness of LLM-generated test cases}\label{sec:codebleu}

LLMs are non-deterministic for constrained inputs, which means that the response to the same input may vary across different executions. In our study, we attempt to use greedy decoding to constrain the response of the LLMs for the same input to produce identical results. We set the temperature to 0, Top K to 1, and Top P to 1. In this section, we analyze whether greedy decoding can ensure consistent results by calculating the CodeBLEU scores of GPT-3.5-turbo generated tests across five different execution times. The evaluation results are presented in \cref{fig:codebleu}. We can observe that the CodeBLEU scores of GPT-3.5-turbo for five different executions are consistently above 85.3\% for each pairwise comparison. However, the scores do not reach 100\% between any two execution times, indicating that there is still some variation in the generated tests despite the use of greedy decoding. 

\subsection{Implications for researchers and developers}

Based on our findings, we present implications for researchers and developers using LLMs for test case generation. Most importantly, our findings indicate that LLM-based testing is more effective at generating tests that protect mature code from regression errors. However, when applied during the early stages of software development on relatively immature code, it is more likely to reinforce existing errors.

Prioritizing correct code and task descriptions is crucial, as our results demonstrate that providing both to LLMs yields the most effective test cases. However, if the correctness of the source code cannot be guaranteed, providing only the task description can still lead to better results than providing incorrect code. 

In addition, it is essential to be cognizant of LLM limitations when working with real-world code, as the effectiveness of LLM-generated test cases is significantly lower in complex, real-world scenarios compared to simpler benchmark datasets (e.g., longer context, function call, and class level tasks), highlighting the need for further research to improve the effectiveness of LLMs in generating test cases for long-context, real-world tasks.

\section{Threat to Validity}

The threat to internal validity lies in the implementation of the empirical study and the analysis of the evaluation results. To reduce the first threat, the authors carefully checked the code twice during the implementation and experiment result analysis stage. To reduce the second threat, the two authors independently analyzed the experiment results and drew experimental conclusions separately. In cases where the conclusions differed, a third, more senior author was consulted to discuss the findings and determine the final result.

The threat to external validity lies in the datasets and the measure tool used in our study. To reduce the threat, we select the three most widely used datasets and two real-world datasets in code generation tasks to measure the effectiveness of LLM-generated test cases. The evaluated subset for each dataset is checked by analyzing whether each task has an incorrect code in all LLM-generated code that can be used for \promptthird. To measure the accuracy of LLM-generated tests, we also use the evaluation tool of HumanEval to ensure the results are correct. Besides, we also use \texttt{coverage.py} to measure the code line coverage of LLM-generated test cases in the correct code, where \texttt{coverage.py} is also widely used by developers and can be relied upon to provide accurate results.

The threat to construction validity lies in the randomness of LLM-generated responses. Since LLMs are non-determinized for their generated response in several different executions with the same input \cite{ouyang2023llm}. To reduce the randomness of LLM-generated responses that would be used to measure the effectiveness of test cases. We use greedy decoding in all of the steps where LLMs would used to generate the response. Moreover, we provided the CodeBLEU results of five different executions of generated tests to demonstrate that our results can reduce the randomness in our experiments, enhancing the overall reliability of our findings.

\section{Conclusion}\label{sec:conclusion}

In this paper, we present the first empirical study on how source code affects the effectiveness of LLM-generated test cases in code generation tasks. Our evaluation results in five open-source and six closed-source models demonstrate that the effectiveness of LLM-generated test cases is highly affected by the prompts used. Providing task descriptions with correct code in the prompt generally leads to higher test case accuracy, better code coverage, and higher bug detection effectiveness compared to other prompts. For example, providing task description and correct code (\promptsecond) achieves 80.4\% test case accuracy in the HumanEval dataset on average for all LLMs at the test level but providing correct code (\promptfour) only achieves 64.1\% accuracy in the HumanEval dataset. Next, we can also observe that \promptsecond also has higher code line coverage compared to other prompts. For example, the average code line coverage for all models of \promptsecond achieves 94.2\% in the APPS dataset for the test level. In contrast, the average code line coverage of other prompts only achieves 92.0\%. Additionally, the bug detection effectiveness of LLM-generated test cases has a similar trend for accuracy and code line coverage. For example, the average bug detection effectiveness of \promptsecond achieves 51.4\% in the APPS dataset, while other prompts only achieve 41.4\% of bug detection effectiveness. We release our source code, datasets, and results in \url{https://anonymous.4open.science/r/ICSE-D15F/}.


\bibliographystyle{ACM-Reference-Format}
\bibliography{sample-base}

\end{document}